\documentstyle[11pt,aaspp]{article}
\newcommand{\onu}{\Omega_\nu}
\newcommand{\oba}{\Omega_{\rm baryon}}
\input epsf.tex
\begin{document}
\title{Cosmological Perturbation Theory in the Synchronous and Conformal
    Newtonian Gauges}
\author{Chung-Pei Ma\footnote{e-mail: cpma@daffy.tapir.caltech.edu}}
\affil{Theoretical Astrophysics 130-33, California Institute of
Technology, Pasadena, CA 91125}
\centerline{and}
\author{Edmund Bertschinger\footnote{e-mail: bertschinger@mit.edu}}
\affil{Department of Physics, Massachusetts Institute of Technology,
Cambridge, MA 02139}
\begin{abstract}
This paper presents a systematic treatment of the linear theory of
scalar gravitational perturbations in the synchronous gauge and the
conformal Newtonian (or longitudinal) gauge.  It differs from others
in the literature in that we give, in both gauges, a complete
discussion of all particle species that are relevant to any flat cold
dark matter (CDM), hot dark matter (HDM), or CDM+HDM models (including
a possible cosmological constant).  The particles considered include
CDM, baryons, photons, massless neutrinos, and massive neutrinos (an
HDM candidate), where the CDM and baryons are treated as fluids while
a detailed phase-space description is given to the photons and
neutrinos.  Particular care is applied to the massive neutrino
component, which has been either ignored or approximated crudely in
previous works.  Isentropic initial conditions on super-horizon scales
are derived.  The coupled, linearized Boltzmann, Einstein and fluid
equations that govern the evolution of the metric and density
perturbations are then solved numerically in both gauges for the
standard CDM model and two CDM+HDM models with neutrino mass densities
$\onu=0.2$ and 0.3, assuming a scale-invariant, adiabatic spectrum of
primordial fluctuations.  We also give the full details of the cosmic
microwave background anisotropy, and present the first accurate
calculations of the angular power spectra in the two CDM+HDM models
including photon polarization, higher neutrino multipole moments, and
helium recombination.  The numerical programs for both gauges are
available at {\tt http://arcturus.mit.edu/cosmics}.
\end{abstract}

\section{Introduction}
The theory of galaxy formation based on gravitational instability is
aimed at describing how primordially-generated fluctuations in matter
and radiation grow into galaxies and clusters of galaxies due to
self-gravity.  A perturbation theory can be formulated when the
amplitudes of the fluctuations are small, and the growth of the
fluctuations can be solved from the linear theory.

In the early universe, gravitational perturbations are inflated to
wavelengths beyond the horizon at the end of the inflationary epoch.
Fluctuations of a given length scale reenter the horizon at a later
time when the horizon has grown to the size of the fluctuations.
Although the process of galaxy formation in recent epochs is well
described by Newtonian gravity (and other physical processes such as
hydrodynamics), a general relativistic treatment is required for
perturbations on scales larger than the horizon size before the
horizon crossing time.  The use of general relativity brought in the
issue of gauge freedom which has caused some confusion over the years.
Lifshitz (1946) adopted the ``synchronous gauge'' for his coordinate
system, which has since become the most commonly used gauge for
cosmological perturbation theories.  However, some complications
associated with this gauge such as the appearance of coordinate
singularities and spurious gauge modes prompted Bardeen (1980) and
others (e.g. Kodama \& Sasaki, 1984) to formulate alternative
approaches that deal only with gauge-invariant quantities.  A thorough
review of the gauge-invariant perturbation theory and its application
to texture-seeded structure formation models is given by Durrer
(1993).  Another possibility is to adopt a different gauge.  We will
discuss in detail in this paper the conformal Newtonian (or the
longitudinal) gauge (Mukhanov, Feldman \& Brandenberger 1992), which
is a particularly convenient gauge to use for scalar perturbations.
It also has the feature that the two scalar fields that describe the
metric perturbations in this gauge are identical (up to a minus sign)
to the gauge-invariant variables Bardeen (1980) constructed.

The linear theory for a perturbed Friedmann-Robertson-Walker universe
was first developed by Lifshitz (1946), later reviewed in Lifshitz \&
Khalatnikov (1963).  The subsequent work can be found summarized in
the textbooks by Weinberg (1972) and Peebles (1980), in the reviews by
Kodama \& Sasaki (1984) and Mukhanov et al (1992),
and in the Summer School lectures by Efstathiou (1990), Bertschinger
(1995), and Bond (1995).  The theory has been applied to various
cosmological models.  In the synchronous gauge, for example, Peebles
\& Yu (1970) solved the linear evolution equations for photons and
baryons in the absence of massless neutrinos and non-baryonic dark
matter.  Bond and Szalay (1983) included the neutrinos but
approximated them as a perfect fluid by using only the first two
moments in the angular expansion of the neutrino distribution
function.  They also did not include CDM.  Xiang \& Kiang (1992)
considered both CDM and massive neutrinos, but treated the latter as
non-relativistic particles and also ignored massless neutrinos.
Holtzman (1989) provided fitting formulas for the baryonic transfer
functions in various cosmological models, but gave no detailed
description of the calculation, and again included only the first two
moments of the neutrino distribution function.  Durrer (1989) obtained
numerical solutions for the perturbations in the gauge-invariant
formalism in the absence of baryons.  Instead of using the convenient
multipole expansion technique, Stompor (1994) adopted the earliest
method by solving the massive neutrino distribution function
directly at discrete momenta and angles.

Another important application of the linear theory is to the cosmic
microwave background anisotropies.  The early calculations of the
radiation perturbations included only the baryons and
the photons (Peebles \& Yu 1970; Wilson \& Silk 1981).
The calculations were subsequently extended to dark matter dominated
models (e.g. Bond \& Efstathiou 1984, 1987; Vittorio \& Silk 1984,
1992; Holtzman 1989).  The post-COBE era has seen a rapid increase in
the number of papers on this subject (see Sugiyama \& Gouda 1992; White,
Scott, \& Silk 1994; and references therein).  However, there has not
been a complete treatment of the massive neutrinos to the accuracy
that is needed for comparison with the near-future anisotropy
experiments.  Truncating the {\it massless} neutrino moments beyond
$l\ge 2$ in the pure CDM model leads to a 10\% error in the anisotropy
power spectrum (Hu et al 1995).  One would expect a comparable or
larger error if the $l\ge 2$ modes were not included for the massless
{\it and} massive neutrinos in CDM+HDM models.

This paper differs from earlier ones in the literature in that we give
a complete discussion of all particle species that are relevant to any
flat CDM, HDM, or CDM+HDM models (including a possible cosmological
constant).  These include CDM, baryons, photons, massless neutrinos,
and massive neutrinos (an HDM candidate).  The recent interests in
CDM+HDM models (e.g. Davis, Summers, \& Schlegel 1992; Klypin et al
1993; Jing et al 1994; Cen \& Ostriker 1994; Ma \& Bertschinger 1994b;
Ma 1994; Primack et al 1995; Klypin et al 1995) prompted us to provide
a detailed treatment of the massive neutrino component, which was
either ignored or crudely approximated in other works.  Part of the
cause for neglecting this component was perhaps a lack of motivation
when the standard CDM model was thought to match the observed
structure well.  Another was perhaps due the sharp increase in the
computing time required to deal with the time-dependent nature of the
massive neutrinos' energy-momentum relation.  For completeness, we
also give full details and results of the calculation of the cosmic
microwave background anisotropy including photon polarization and
helium recombination.

This paper serves two purposes.  First, it is an independent paper in
which we present a systematic treatment of the linear theory of scalar
isentropic gravitational perturbations in the synchronous and
conformal Newtonian gauges.  The coupled, linearized Einstein,
Boltzmann, and fluid equations for the metric and density
perturbations are presented in parallel in the two gauges.  The CDM
and the baryon components behave like collisionless and collisional
fluids, respectively, while the photons and the neutrinos require a
phase-space description governed by the Boltzmann transport equation.
We also derive analytically the time dependence of the perturbations
on scales larger than the horizon to illustrate the dependence on the
gauge choice.  This information is needed in the initial conditions
for the numerical integration of the evolution equations.

This paper also serves as a companion paper to Ma \& Bertschinger
(1994a) in which we reported the main results from our linear
calculation of the full neutrino phase space in a CDM+HDM model with
$\Omega_{\rm c}=0.65$, $\onu=0.3$, $\oba=0.05$, and $H_0=50$ km
s$^{-1}$ Mpc$^{-1}$.  (The corresponding neutrino mass is $m_\nu = 7$
eV.)  The motivation was to obtain an accurate sampling of the
neutrino phase space for the HDM initial conditions in subsequent
$N$-body simulations (Ma \& Bertschinger 1994b; Ma 1994) of structure
formation in CDM+HDM models.  We adopted a two-step Monte Carlo
procedure to achieve this goal: (1) Integrate the coupled, linearized
Boltzmann, Einstein, and fluid equations for all particle species in
the model (i.e., CDM, HDM, photons, baryons and massless neutrinos) to
obtain the evolution of the metric perturbations; (2) Follow the
trajectories of individual neutrinos by integrating the geodesic
equations using the metric computed in (1).  Since no coordinate
singularities occur in the conformal Newtonian gauge and the geodesic
equations have simple forms, the geodesic integration in step (2) was
carried out in this gauge, starting shortly after neutrino decoupling
at redshift $z\sim 10^9$ until $z = 13.5$.  We focus on step (1) in
this paper.  Following historical precedents, we first developed the
code for the Boltzmann integration in the synchronous gauge.  The
transformation relating the synchronous gauge and the conformal
Newtonian gauge was then derived and used to compute the metric
perturbations in the latter gauge for step (2) of the calculation.
Subsequently we developed a code to perform the full integration in
the conformal Newtonian gauge.

The organization of this paper is as follows.  In \S 2 we write down
the metric for the two gauges and summarize their properties.  In \S
3, we derive the gauge transformation relating two arbitrary gauges
and obtain the transformation between the synchronous and the
conformal Newtonian gauges.  The linearized evolution equations for
the metric and the density perturbations are given in \S\S 4 and 5.
Section 4 discusses the Einstein equations with emphasis on the source
terms, the energy-momentum tensor, in the two gauges.  The perturbed
fluid equations are derived from the energy-momentum conservation,
which are applied to CDM and the baryons in \S 5.  The rest of \S 5
contains detailed treatments of the photon and neutrino phase space
distributions, recombination, and the coupling of photons and baryons.
The photon and neutrino distribution functions are expanded in
Legendre polynomials, reducing the linearized Boltzmann equation to a
set of coupled ordinary differential equations for the expansion
modes.  The massive neutrinos require a slightly more complicated
treatment because the energy-momentum relation depends nontrivially on
time when the neutrinos make a transition from being relativistic to
non-relativistic.  Our method for computing the microwave background
anisotropy is presented in \S 6.  Section 7 discusses the behavior of
the perturbations before horizon crossing.  The necessary initial
conditions for the variables in the two gauges are given.  Section 8
presents the numerical results for the evolution of the perturbations
and the angular power spectrum of the microwave anisotropy in CDM+HDM
models.

  The complex physics we study requires the use of many
equations and symbols.  As a guide to the reader, in Table 1 we
summarize the key symbols and the equations where they are defined
or first used.

\begin{table}
\begin{tabular}{|c|l|c|}  \hline
Symbol    &	Meaning 	&	Reference Eqn\\ \hline\hline
$a$	  & 	scale factor 	&	(1)	\\ \hline
$\tau$	  & 	conformal time  &	(1)	\\ \hline
$k$	  & 	wavenumber of Fourier mode & (\ref{hijk})	\\ \hline
$P_i$	  & 	conjugate momentum to (comoving) position $x^i$ &
		(\ref{psubi}) \\ \hline
$p_i$	  & 	proper momentum & (\ref{psubi})	\\ \hline
$q_i$	  & 	$=ap_i$ 	& --	\\ \hline
$\epsilon$  & 	$=(q^2+a^2 m^2)^{1/2}$ &  --	\\ \hline
$h$       & 	synchronous &	(1), (\ref{hijk})\\
$\eta$    & 	metric perturbations &  \\ \hline
$\phi$    & 	conformal Newtonian  &	(\ref{conformal})\\
$\psi$    &	metric perturbations &	\\ \hline
subscript $c$ 	&	cold dark matter & -- \\ \hline
subscript $\nu$ &	massless neutrinos & -- \\ \hline
subscript $h$	& 	massive neutrinos & -- \\ \hline
subscript $\gamma$	& 	photons & -- \\ \hline
subscript $b$	& 	baryons & -- \\ \hline
$f_0$	  &   	unperturbed phase space distribution function
	  & (\ref{equil-dist})\\ \hline
$\Psi$	  &   	perturbation to $f_0$ & (\ref{f-pert})\\ \hline
$\Psi_l$  &   	$l$th Legendre component of $\Psi$ & (\ref{psisubl})\\ \hline
$F_l$	  &   	momentum-averaged $\Psi_l$  & (\ref{fsubl}) \\ \hline
$G_{\gamma\,l}$&photon polarization component  & (\ref{thom2}) \\ \hline
$\delta$  &	density fluctuation $(=F_0)$ &
		(\ref{deltat}a), (\ref{fsubnu}) \\ \hline
$\theta$  &	divergence of fluid velocity $(=3kF_1/4)$ &
		(\ref{theta}), (\ref{deltat}b), (\ref{fsubnu}) \\ \hline
$\sigma$  & 	shear stress  $(=F_2/2)$  &
		(\ref{theta}), (\ref{deltat}d), (\ref{fsubnu}) \\ \hline
$\Delta$  &	temperature fluctuation $(=\Delta T/T=F_\gamma/4)$ &
		(\ref{Delta}) \\ \hline
$\Delta_l$ &    $l$th Legendre mode of $\Delta$ $(=F_{\gamma\,l}/4)$ &
		(\ref{Delta-exp1}) \\ \hline
$C_l$	  &	temperature fluctuation power spectrum &
		(\ref{C_l-2}) \\ \hline
$c_s$	& 	baryon sound speed ($=\delta P_b/\delta\rho_b$) &
		(\ref{soundsp})\\ \hline
$c_{pb}$& 	sound speed of coupled photon-baryon fluid  &
		(\ref{nostress})\\ \hline
$w$	& 	describes equation of state ($=P/\rho$) &
		(\ref{fluid}),(\ref{fluid2}) \\ \hline
$n_e$   &	electron number density & (\ref{scatop})\\ \hline
$\sigma_T$   &	Thomson scattering cross section & (\ref{scatop})\\ \hline
$\tau_c$     &  $=(a n_e \sigma_T)^{-1}$ & (\ref{thetabdot}) \\ \hline
$R_\nu$	     &  density ratio $\bar\rho_\nu/(\bar\rho_\gamma+\bar\rho_\nu)$ &
		(\ref{press}) \\ \hline
$R$	     &  density ratio $(4/3)\bar\rho_\gamma/\bar\rho_b$ &
		(\ref{momentum}) \\ \hline
\end{tabular}
\caption{Symbols used in this paper.}
\end{table}

\section{The Two Gauges}
We consider only spatially flat ($\Omega=1$) background spacetimes
with isentropic scalar metric perturbations.  The spacetime
coordinates are denoted by $x^\mu$, $\mu\in(0,1,2,3)$, where $x^0$ is
the time component and $x^i$, $i\in(1,2,3)$ are the spatial components
in Cartesian coordinates.  Greek letters $\alpha,\beta,\gamma$ and so
on always run from 0 to 3, labeling the four spacetime-coordinates;
Roman letters such as $i, j, k$ always run from 1 to 3, labeling the
spatial parts of a four-vector.  Repeated indices are summed.  Since
our interests lie in the physics in an expanding universe, we use
comoving coordinates $x^\mu = (\tau,\vec{x})$ with the expansion
factor $a(\tau)$ of the universe factored out.  The comoving
coordinates are related to the proper time and positions $t$ and
$\vec{r}$ by $dx^0 =d\tau = dt/a(\tau)$, $d\vec{x}=d\vec{r}/a(\tau)$.
Dots will denote derivatives with respect to $\tau$: $\dot
a\equiv\partial a/\partial\tau$.  The speed of light $c$ is set to
unity.

The components $g_{00}$ and $g_{0i}$ of the metric tensor in the
synchronous gauge are by definition unperturbed.  The line element is
given by
\begin{equation}
  ds^2 = a^2(\tau)\{-d\tau^2 + (\delta_{ij} + h_{ij})dx^i dx^j\}\,.
\end{equation}
The metric perturbation $h_{ij}$ can be decomposed into a trace part
$h \equiv h_{ii}$ and a traceless part consisting of three
pieces, $h^\parallel_{ij}, h^\perp_{ij}$, and $h^T_{ij}$, where
$h_{ij}=h\delta_{ij}/3 + h^\parallel_{ij}+h^\perp_{ij}+h^T_{ij}$.
By definition, the divergences of $h^\parallel_{ij}$ and
$h^\perp_{ij}$ (which are vectors) are longitudinal and transverse,
respectively, and $h^T_{ij}$ is transverse, satisfying
\begin{equation}
\label{decomp}
     \epsilon_{ijk} \partial_j \partial_l h^\parallel_{lk} = 0\,,
	\qquad \partial_i\partial_j h^\perp_{ij} = 0\,,\qquad
     \partial_i h^T_{ij} = 0 \,.
\end{equation}
It then follows that $h^\parallel_{ij}$ can be written in terms
of some scalar field $\mu$ and $h^\perp_{ij}$ in terms of
some divergenceless vector $\vec{A}$ as
\begin{eqnarray}
        h^\parallel_{ij} &=& \left( \partial_i\partial_j -
	{1\over 3} \delta_{ij} \nabla^2 \right) \mu\,, \nonumber\\
        h^\perp_{ij} &=& \partial_i A_j + \partial_j A_i\,,\qquad
        \partial_i A_i = 0\,.
\end{eqnarray}
The two scalar fields $h$ and $\mu$ (or $h^\parallel_{ij}$)
characterize the scalar mode of the metric perturbations,
while $A_i$ (or $h^\perp_{ij}$) and $h^T_{ij}$ represent the vector
and the tensor modes, respectively.

We will be working in the Fourier space $k$ in this paper.
We introduce two fields $h(\vec{k},\tau)$ and $\eta(\vec{k},\tau)$ in
$k$-space and write the scalar mode of $h_{ij}$ as a Fourier integral
\begin{equation}
\label{hijk}
	h_{ij}(\vec{x},\tau) = \int d^3k e^{i\vec{k}\cdot\vec{x}}
	\left\{ \hat{k}_i\hat{k}_j h(\vec{k},\tau) +
	(\hat{k}_i\hat{k}_j - {1 \over 3}\delta_{ij})\,
        6\eta(\vec{k},\tau) \right\} \,,\quad \vec{k} = k\hat{k} \,.
\end{equation}
Note that $h$ is used to denote the trace of $h_{ij}$ in both the
real space and the Fourier space.

In spite of its wide-spread use, there are serious disadvantages
associated with the synchronous gauge.  Since the choice of the
initial hypersurface and its coordinate assignments are arbitrary,
the synchronous gauge conditions do not fix the gauge degrees of
freedom completely.  Such residual gauge freedom is manifested in the
spurious gauge modes contained in the solutions to the equations
for the density perturbations.  The appearance of these modes has
caused some confusion over the years and prompted Bardeen (1980)
to formulate alternative
approaches that deal only with gauge-invariant quantities.  Another
difficulty with the synchronous gauge is that since the coordinates
are defined by freely falling observers, coordinate singularities
arise when two observers' trajectories intersect each other: a point
in spacetime will have two coordinate labels.  A different initial
hypersurface of constant time has to be chosen to remove these
singularities.

The conformal Newtonian gauge (also known as the longitudinal gauge)
advocated by Mukhanov et al. (1992) is a particularly simple gauge to use
for the scalar mode of metric perturbations.  The perturbations are
characterized by two scalar potentials $\psi$ and $\phi$ which appear
in the line element as
\begin{equation}
\label{conformal}
    ds^2 = a^2(\tau)\left\{ -(1+2\psi)d\tau^2 +
        (1-2\phi)dx^i dx_i \right\} \,.
\end{equation}
It should be emphasized that the conformal Newtonian gauge is a
restricted gauge since the metric is applicable only for the scalar
mode of the metric perturbations; the vector and the tensor degrees of
freedom are eliminated from the beginning.  Nonetheless, it can be
easily generalized to include the vector and the tensor modes
(Bertschinger 1995).  The discussion here will be confined to the
scalar perturbations only.

One advantage of working in this gauge is that the metric tensor
$g_{\mu\nu}$ is diagonal.  This simplifies the calculations and leads
to simple geodesic equations (Ma \& Bertschinger 1994a).  Another
advantage is that $\psi$ plays the role of the gravitational potential
in the Newtonian limit and thus has a simple physical interpretation.
Moreover, the two scalar potentials $\psi$ and $\phi$ in this gauge
are related to the gauge-invariant variables $\Phi_A$ and $\Phi_H$ of
Bardeen (1980) and $\Psi$ and $\Phi$ of Kodama \& Sasaki (1984) by the
simple relation
\begin{equation}
	\psi=\Phi_A=\Psi\,,\qquad \phi=-\Phi_H=-\Phi\,.
\end{equation}
No gauge modes are present in this gauge to obscure the meaning of the
physical modes since the gauge freedom is completely fixed for
$\Omega=1$ aside from the addition of spatial constants to $\psi$ and
$\phi$.  The two potentials differ when the energy-momentum tensor
$T^\mu{}_{\!\nu}$ contains a nonvanishing traceless and longitudinal
component (see eq.~[\ref{ein-cond}]).

\section{Gauge Transformations}
In this section we first derive the transformation law relating two
arbitrary gauges.  From it, the gauge transformation relating the
synchronous gauge and the conformal Newtonian gauge is readily obtained.

A perturbed flat Friedmann-Robertson-Walker metric can be written in
general as
\begin{eqnarray}
\label{perturb}
  g_{00} &=& -a^2(\tau)\,\left\{ 1+2\psi(\vec{x},\tau) \right\}\,,
	\nonumber\\
  g_{0i} &=& a^2(\tau)\,w_i(\vec{x},\tau)\,,\\
  g_{ij} &=& a^2(\tau)\,\left\{ [1-2\phi(\vec{x},\tau)] \delta_{ij}
        + \chi_{ij}(\vec{x},\tau) \right\}\,,\qquad \chi_{ii}=0
	\nonumber
\end{eqnarray}
where the functions $\psi, \phi, w_i$ and $\chi_{ij}$ represent
metric perturbations about the Robertson-Walker spacetime
and are assumed to be small compared with unity.  The trace part of
the perturbation to $g_{ij}$ is absorbed in $\phi$, and $\chi_{ij}$
is taken to be traceless.

Consider a general coordinate transformation from a coordinate
system $x^\mu$ to another $\hat{x}^\mu$
\begin{equation}
        x^\mu \rightarrow \hat{x}^\mu = x^\mu + d^\mu(x^\nu)\,.
\end{equation}
We write the time and the spatial parts separately as
\begin{eqnarray}
\label{shift}
        \hat{x}^0 &=& x^0 + \alpha(\vec{x},\tau)\,, \nonumber\\
        \hat{\vec{x}} &=& \vec{x} + \vec{\nabla}\beta(\vec{x},\tau)
                + \vec{\epsilon}\,(\vec{x},\tau)\,,\quad
                \vec{\nabla}\cdot\vec{\epsilon} = 0\,,
\end{eqnarray}
where the vector $\vec{d}$ has been decomposed into a longitudinal
component $\vec{\nabla}\beta\,$
($\vec{\nabla}\times\vec{\nabla}\beta=0$)
and a transverse component
$\vec{\epsilon}\,$ ($\vec{\nabla}\cdot\vec{\epsilon}=0$).  The
requirement that $ds^2$ be invariant under this coordinate
transformation leads to
\begin{equation}
        \hat{g}_{\mu\nu}(x) = g_{\mu\nu}(x)
        - g_{\mu\beta}(x) \partial_\nu d^\beta
        - g_{\alpha\nu}(x) \partial_\mu d^\alpha
        - d^\alpha \partial_\alpha g_{\mu\nu}(x) + O(d^2) \,.
\end{equation}
We note that both sides of this equation are evaluated at the same
coordinate values $x$ in the two gauges, which do not correspond to
the same physical point in general.  Assuming $d^\mu$ to be of the
same order as the metric perturbations $\psi,w_i,\phi$ and
$\chi_{ij}$, the metric perturbations in the two coordinate systems
are related to first order in the perturbed quantities by
\begin{mathletters}
\begin{eqnarray}
    \hat{\psi}(\vec{x},\tau) &=&
       \psi (\vec{x},\tau) - \dot{\alpha}(\vec{x},\tau)
       - {\dot{a}\over a} \alpha(\vec{x},\tau)\,,
	\label{trans1a} \\
    \hat{w}_i(\vec{x},\tau) &=&
          w_i(\vec{x},\tau) + \partial_i \alpha(\vec{x},\tau)
          - \partial_i \dot{\beta}(\vec{x},\tau)
          - \dot{\epsilon}_i(\vec{x},\tau)\,,
	\label{trans1b} \\
    \hat{\phi}(\vec{x},\tau) &=&
          \phi (\vec{x},\tau) + {1 \over 3}\nabla^2\beta(\vec{x},\tau)
          + {\dot{a} \over a} \alpha(\vec{x},\tau)\,,
	\label{trans1c} \\
    \hat{\chi}_{ij}(\vec{x},\tau) &=&
          \chi_{ij}(\vec{x},\tau) - 2\left\{ \left(
        \partial_i\partial_j - {1\over 3}\delta_{ij}\nabla^2 \right)
        \beta(\vec{x},\tau) + {1\over 2}\left( \partial_i\epsilon_j +
        \partial_j\epsilon_i \right) \right\}\,. \label{trans1d}
\end{eqnarray}
\label{trans1}
\end{mathletters}
We can further decompose the transformations of $w_i$ and $\chi_{ij}$
above into longitudinal and transverse parts:
\begin{eqnarray}
\label{w}
    \hat{w}^\parallel_i(\vec{x},\tau) &=&
          w^\parallel_i(\vec{x},\tau) + \partial_i \alpha(\vec{x},\tau)
          - \partial_i \dot{\beta}(\vec{x},\tau)\,, \nonumber\\
    \hat{w}^\perp_i(\vec{x},\tau) &=&
          w^\perp_i(\vec{x},\tau) - \dot{\epsilon}_i(\vec{x},\tau)\,,
\end{eqnarray}
and
\begin{eqnarray}
\label{chi}
    \hat{\chi}^\parallel_{ij}(\vec{x},\tau) &=&
        \chi^\parallel_{ij}(\vec{x},\tau) - 2\left(\partial_i\partial_j
        - {1\over 3}\delta_{ij}\nabla^2 \right) \beta(\vec{x},\tau)\,,
        \nonumber\\
    \hat{\chi}^\perp_{ij}(\vec{x},\tau) &=&
          \chi^\perp_{ij}(\vec{x},\tau) -
          (\partial_i\epsilon_j + \partial_j\epsilon_i)\,, \nonumber\\
    \hat{\chi}^T_{ij}(\vec{x},\tau) &=& \chi^T_{ij}(\vec{x},\tau) \,,
\end{eqnarray}
where $w_i = w^\parallel_i + w^\perp_i$, $\chi_{ij} =
\chi^\parallel_{ij}+\chi^\perp_{ij}+\chi^T_{ij}\,$,
and $\chi^\parallel_{ij}\,$, $\chi^\perp_{ij}$ and $\chi^T_{ij}$
obey equation (\ref{decomp}).  Equations (\ref{trans1})--(\ref{chi})
describe the transformation of metric perturbations under a general
infinitesimal coordinate transformation.

We can now use equations (\ref{trans1}) to relate the scalar metric
perturbations $(\phi, \psi)$ in the conformal Newtonian gauge to $h_{ij}
=h\delta_{ij}/3+h^\parallel_{ij}$ in the synchronous gauge.  Let $
\hat{x}^\mu$ denote the synchronous coordinates and $x^\mu$ the
conformal Newtonian coordinates with $\hat{x}^\mu=x^\mu+d^\mu$.  From
equations (\ref{w}) and (\ref{chi}), we find
\begin{mathletters}
\begin{eqnarray}
       \alpha(\vec{x},\tau) &=& \dot{\beta}(\vec{x},\tau) + \xi(\tau)
		 \label{consyn1a} \,,\label{alpha}\\
       \epsilon_i(\vec{x},\tau) &=& \epsilon_i(\vec{x})
		 \label{consyn1b} \,,\\
       h^\parallel_{ij}(\vec{x},\tau) &=& -2\left(
            \partial_i\partial_j - {1\over 3}\delta_{ij}\nabla^2 \right)
            \beta(\vec{x},\tau)  \label{consyn1c} \,,\\
       \partial_i\epsilon_j + \partial_j\epsilon_i &=& 0\,,
		\label{consyn1d}
\end{eqnarray}
\end{mathletters}
where $\xi(\tau)$ is an arbitrary function of time, reflecting the
gauge freedom associated with the coordinate transformation:
$\hat{x}^0 = x^0 + \xi(\tau)$, $\hat{x}^i = x^i$.
This transformation corresponds to a global redefinition of the units
of time with no physical significance; therefore we shall set $\xi=0$
from now on.  From equations (\ref{trans1a}) and (\ref{trans1c}) we then
obtain
\begin{eqnarray}
\label{consyn2}
       \psi(\vec{x},\tau) &=&
            +\ddot{\beta}(\vec{x},\tau) +
            {\dot{a}\over a}\dot{\beta}(\vec{x},\tau)\,,\nonumber\\
       \phi(\vec{x},\tau) &=& -{1\over 6} h(\vec{x},\tau) -
	    {1 \over 3}\nabla^2\beta(\vec{x},\tau)
	    - {\dot{a}\over a} \dot{\beta}(\vec{x},\tau)\,,
\end{eqnarray}
and $\beta$ is determined by $h^\parallel$ in equation (\ref{consyn1c}).

In terms of $h$ and $\eta$ introduced in equation (\ref{hijk}),
$h^\parallel_{ij}$ in the synchronous gauge is given by
\begin{equation}
\label{hpara}
        h^\parallel_{ij}(\vec{x},\tau) = \int d^3k
           \,e^{i\vec{k}\cdot\vec{x}}\,(\hat{k}_i\hat{k}_j -
	   {1 \over 3}\delta_{ij})
	   \left\{ h(\vec{k},\tau) + 6\eta(\vec{k},\tau) \right\}\,,
	   \quad \vec{k} = k\hat{k} \,.
\end{equation}
Comparing $h^\parallel_{ij}$ in equations (\ref{consyn1c}) and
(\ref{hpara}), we can read off $\beta$:
\begin{equation}
\label{beta}
    \beta(\vec{x},\tau) = \int d^3k\,e^{i\vec{k}\cdot\vec{x}}
        \,{1\over 2k^2} \left\{ h(\vec{k},\tau) + 6\eta(\vec{k},\tau)
	\right\} \,.
\end{equation}
Then from equations (\ref{consyn2}), the conformal Newtonian potentials
$\phi$ and $\psi$ are related to the synchronous potentials $h$ and
$\eta$ in $k$-space by
\begin{eqnarray}
\label{trans2}
     \psi(\vec{k},\tau) &=& {1\over 2k^2} \left\{\ddot{h}(\vec{k},\tau)
	+ 6\ddot{\eta}(\vec{k},\tau) + {\dot{a} \over a} \left[
	\dot{h}(\vec{k},\tau) + 6\dot{\eta}(\vec{k},\tau)
        \right] \right\} \,,\nonumber\\
     \phi(\vec{k},\tau) &=&
         \eta(\vec{k},\tau) - {1\over 2k^2}{\dot{a} \over a}
         \left[ \dot{h}(\vec{k},\tau) + 6\dot{\eta}(\vec{k},\tau) \right] \,.
\end{eqnarray}
The other components of the metric perturbations, $w_i,
\chi^\perp_{ij}\,$, and $\chi^T_{ij}\,$, are zero in both gauges.

\section{Einstein Equations and Energy-Momentum Conservation}
For a homogeneous Friedmann-Robertson-Walker universe with energy
density $\bar{\rho}(\tau)$ and pressure $\bar{P}(\tau)$, the Einstein
equations give the following evolution equations for the expansion
factor $a(\tau)$:
\begin{eqnarray}
\label{friedmann}
	\left( {\dot{a}\over a} \right)^2 &=& {8\pi\over
		3}Ga^2 \bar{\rho} - \kappa \,,\\
	{d\over d\tau} \left( {\dot{a}\over a} \right)
		&=& -{4\pi\over 3}Ga^2 (\bar{\rho}+3\bar{P}) \,,
\label{friedmann2}
\end{eqnarray}
where the dots denote derivatives with respect to $\tau$, and $\kappa$
is positive, zero, or negative for a closed, flat, or open universe,
respectively.  We consider only models with total $\Omega=1$ in this
paper, so we set $\kappa=0$.  A cosmological constant is allowed through
its inclusion in $\bar\rho$ and $\bar P$: $\bar\rho_\Lambda=\Lambda/8\pi G
=-\bar P_\Lambda$.  This is the only place that $\Lambda$ enters in the
entire set of calculations.  It follows from equation (\ref{friedmann})
with $\kappa=0$ that the expansion factor scales as $a\propto\tau$ in
the radiation-dominated era, $a\propto\tau^2$ in the matter-dominated era,
and $a\propto(\tau_\infty-\tau)^{-1}$ in a cosmological constant-dominated
era (in the latter case, $\tau_\infty$ is the radius of the de Sitter
event horizon).

We find it most convenient to solve the linearized Einstein equations
in the two gauges in the Fourier space $k$.  In the synchronous gauge,
the scalar perturbations are characterized by $h(\vec{k},\tau)$ and
$\eta(\vec{k},\tau)$ in equation (\ref{hijk}).  In terms of $h$ and
$\eta$, the time-time, longitudinal time-space, trace space-space, and
longitudinal traceless space-space parts of the Einstein equations
give the following four equations to linear order in $k$-space:
\medskip
\newline{\it Synchronous gauge ---\hfil}
\begin{mathletters}
\begin{eqnarray}
    k^2\eta - {1\over 2}{\dot{a}\over a} \dot{h}
        &=& 4\pi G a^2 \delta T^0{}_{\!0}(\mbox{Syn})
	\,,\label{ein-syna}\\
    k^2 \dot{\eta} &=& 4\pi Ga^2 (\bar{\rho}+\bar{P})
	\theta(\mbox{Syn})	 \,,\label{ein-synb}\\
    \ddot{h} + 2{\dot{a}\over a} \dot{h} - 2k^2
	\eta &=& -8\pi G a^2 \delta T^i{}_{\!i}(\mbox{Syn})
	\,,\label{ein-sync}\\
    \ddot{h}+6\ddot{\eta} + 2{\dot{a}\over a}\left(\dot{h}+6\dot{\eta}
	\right) - 2k^2\eta &=& -24\pi G a^2 (\bar{\rho}+\bar{P})
	\sigma(\mbox{Syn})	\,.\label{ein-synd}
\end{eqnarray}
\label{ein-syn}
\end{mathletters}
The label ``Syn'' is used to distinguish the components of the
energy-momentum tensor in the synchronous gauge from those in the
conformal Newtonian gauge.  The variables $\theta$ and $\sigma$ are
defined as
\begin{equation}
\label{theta}
 	(\bar{\rho}+\bar{P})\theta \equiv i k^j \delta T^0{}_{\!j}\,,
	\qquad	(\bar{\rho}+\bar{P})\sigma \equiv -(\hat{k}_i\hat{k}_j
	- {1\over 3} \delta_{ij})\Sigma^i{}_{\!j}\,,
\end{equation}
and $\Sigma^i{}_{\!j} \equiv T^i{}_{\!j}-\delta^i{}_{\!j} T^k{}_{\!k}/3$
denotes the traceless component of $T^i{}_{\!j}$.
Kodama \& Sasaki (1984) define the anisotropic stress perturbation $\Pi$,
related to our $\sigma$ by $\sigma=2\Pi\bar P/3(\bar\rho+\bar P)$.
When the different components of matter and radiation (i.e., CDM,
HDM, baryons, photons, and massless neutrinos) are treated
separately, $(\bar{\rho}+\bar{P})\theta = \sum_i
(\bar{\rho}_i+\bar{P}_i)\theta_i$ and $(\bar{\rho}+\bar{P})\sigma =
\sum_i (\bar{\rho}_i+\bar{P}_i)\sigma_i\,$, where the index $i$ runs
over the particle species.

In the conformal Newtonian gauge, the first-order perturbed Einstein
equations give
\medskip
\newline{\it Conformal Newtonian gauge ---\hfil}
\begin{mathletters}
\begin{eqnarray}
    k^2\phi + 3{\dot{a}\over a} \left( \dot{\phi} + {\dot{a}\over a}\psi
	\right) &=& 4\pi G a^2 \delta T^0{}_{\!0}(\mbox{Con}) \,,
	\label{ein-cona}\\
    k^2 \left( \dot{\phi} + {\dot{a}\over a}\psi \right)
	 &=& 4\pi G a^2 (\bar{\rho}+\bar{P}) \theta(\mbox{Con})
	 \,,\label{ein-conb}\\
    \ddot{\phi} + {\dot{a}\over a} (\dot{\psi}+2\dot{\phi})
	+\left(2{\ddot{a} \over a} - {\dot{a}^2 \over a^2}\right)\psi
	+ {k^2 \over 3} (\phi-\psi)
	&=& {4\pi\over 3} G a^2 \delta T^i{}_{\!i}(\mbox{Con})
	\,,\label{ein-conc}\\
    k^2(\phi-\psi) &=& 12\pi G a^2 (\bar{\rho}+\bar{P})\sigma(\mbox{Con})
	\,,\label{ein-cond}
\end{eqnarray}
\label{ein-con}
\end{mathletters}
where ``Con'' labels the conformal Newtonian coordinates.

Now we derive the transformation relating $\delta T^\mu{}_{\!\nu}$ in
the two gauges.  For a perfect fluid of energy density $\rho$ and
pressure $P$, the energy-momentum tensor has the form
\begin{equation}
	T^\mu{}_{\!\nu} = P g^\mu{}_{\!\nu} + (\rho+P) U^\mu U_\nu \,,
\end{equation}
where $U^\mu = dx^\mu /\sqrt{-ds^2} $ is the four-velocity of the
fluid.  The pressure $P$ and energy density $\rho$ of a perfect fluid
at a given point are defined to be the pressure and energy density
measured by a comoving observer at rest with the fluid at the instant
of measurements.  For a fluid moving with a small coordinate velocity
$v^i \equiv dx^i/d\tau$, $v^i$ can be treated as a perturbation of the
same order as $\delta\rho=\rho-\bar{\rho}$, $\delta P=P-\bar{P}$, and
the metric perturbations.  Then to linear order in the perturbations the
energy-momentum tensor is given by
\begin{eqnarray}
	T^0{}_{\!0} &=& -(\bar{\rho} + \delta\rho) \,,\nonumber\\
	T^0{}_{\!i} &=& (\bar{\rho}+\bar{P}) v_i  = -T^i{}_{\!0}\,,\nonumber\\
	T^i{}_{\!j} &=& (\bar{P} + \delta P) \delta^i{}_{\!j}
		+ \Sigma^i{}_{\!j} \,,\qquad
		\Sigma^i{}_{\!i}=0 \,,
\end{eqnarray}
where we have allowed an anisotropic shear perturbation
$\Sigma^i{}_{\!j}$ in $T^i{}_{\!j}$.  As we shall see, since the
photons are tightly coupled to the baryons before recombination, the
dominant contribution to this shear stress comes from the neutrinos.
We note that for a fluid, $\theta$ defined in equation (\ref{theta}) is
simply the divergence of the fluid velocity: $\theta = ik^j v_j$.

The energy-momentum tensor $T^\mu{}_{\!\nu}(\mbox{Syn})$ in
the synchronous gauge is related to $T^\mu{}_{\!\nu}(\mbox{Con})$ in
the conformal Newtonian gauge by the transformation
\begin{equation}
        T^\mu{}_{\!\nu}(\mbox{Syn}) =
	{\partial \hat{x}^\mu \over \partial x^\sigma}
        {\partial x^\rho \over \partial \hat{x}^\nu}
	T^\sigma{}_{\!\rho}(\mbox{Con})\,,
\end{equation}
where $\hat{x}^\mu$ and $x^\mu$ denote the synchronous and the
conformal Newtonian coordinates, respectively.  It follows that to
linear order, $T^0{}_{\!0}(\mbox{Syn})= T^0{}_{\!0}(\mbox{Con})\,,
T^0{}_{\!j}(\mbox{Syn})= T^0{}_{\!j}(\mbox{Con})+ik_j\alpha
(\bar{\rho}+\bar{P})\,$, and
$T^i{}_{\!j}(\mbox{Syn})=T^i{}_{\!j}(\mbox{Con})\,$, where $\alpha =
\hat{x}^0 - x^0 = (\dot{h}+6\dot{\eta})/2k^2$ in $k$-space from
equations (\ref{alpha}) and (\ref{beta}).  Let $\delta \equiv
\delta\rho/\bar{\rho}=-\delta T^0{}_{\!0}/\bar{\rho}$.  Evaluating
the perturbations at the same spacetime coordinate values, we obtain
\begin{mathletters}
\begin{eqnarray}
	\delta(\mbox{Syn}) &=& \delta(\mbox{Con})
		- \alpha {\dot{\bar\rho} \over {\bar\rho}}\,,\\
	\theta(\mbox{Syn}) &=& \theta(\mbox{Con}) - \alpha k^2\,,\\
	\delta P(\mbox{Syn}) &=& \delta P(\mbox{Con})
		-\alpha\dot{\bar{P}} \,,\\
	\sigma(\mbox{Syn}) &=& \sigma(\mbox{Con}) \,.
\end{eqnarray}
\label{deltat}
\end{mathletters}
This transformation also applies to individual species when more than one
particle species contributes to the energy-momentum tensor, provided that
the appropriate $\bar{\rho}$ and $\bar{P}$ are used for each component.

The non-relativistic fluid description is appropriate for the CDM and
the baryon components.  The photon and the neutrino components,
however, can be appropriately described only by their full
distribution functions in phase space.  The energy-momentum tensor in
this case is expressed through integrals over momenta of the
distribution functions.  We will discuss it in detail in \S 5.

The conservation of energy-momentum is a consequence of the Einstein
equations.  Let $w \equiv P/\rho$ describe the equation of state.
Then the perturbed part of energy-momentum conservation equations
\begin{equation}
\label{Econs}
	T^{\mu\nu}{}_{\!;\mu} = \partial_\mu T^{\mu\nu}
	+ \Gamma^\nu{}_{\!\alpha\beta} T^{\alpha\beta}
	+ \Gamma^\alpha{}_{\!\alpha\beta} T^{\nu\beta} = 0
\end{equation}
in $k$-space implies
\medskip
\newline{\it Synchronous gauge ---\hfil}
\begin{eqnarray}
\label{fluid}
	\dot{\delta} &=& - (1+w) \left(\theta+{\dot{h}\over 2}\right)
	  - 3{\dot{a}\over a} \left({\delta P \over \delta\rho} - w
	  \right)\delta  \,,\nonumber\\
	\dot{\theta} &=& - {\dot{a}\over a} (1-3w)\theta - {\dot{w}\over
	     1+w}\theta + {\delta P/\delta\rho \over 1+w}\,k^2\delta
	     - k^2 \sigma\,,
\end{eqnarray}
\newline{\it Conformal Newtonian gauge ---\hfil}
\begin{eqnarray}
\label{fluid2}
	\dot{\delta} &=& - (1+w) \left(\theta-3{\dot{\phi}}\right)
	  - 3{\dot{a}\over a} \left({\delta P \over \delta\rho} - w
	  \right)\delta \,,\nonumber\\
	\dot{\theta} &=& - {\dot{a}\over a} (1-3w)\theta - {\dot{w}\over
	     1+w}\theta + {\delta P/\delta\rho \over 1+w}\,k^2\delta
	     - k^2 \sigma + k^2 \psi \,.
\end{eqnarray}
These equations are valid for a single uncoupled fluid, or for the net
(mass-averaged) $\delta$ and $\theta$ for all fluids.  They need to be
modified for individual components if the components interact with each
other.  An example is the baryonic fluid in our model, which couples
to the photons before recombination via Thomson scattering.
In the next section we will show that an extra term representing
momentum transfer between the two components needs to be added to the
$\dot\delta$ equation for the baryons.

For the isentropic primordial perturbations considered in this paper,
the equations above simplify since $\delta P=c_s^2\delta\rho$,
where $c_s^2=dP/d\rho=w+\rho dw/d\rho$ is the adiabatic sound speed
squared.  For the photons and baryons (the only collisional fluid
components with pressure), $w$ is a constant ($w=1/3$ for photons and
$w\approx0$ for baryons since they are nonrelativistic at the times of
interest).  Thus, $\delta P/\delta\rho-w=0$.  The entropy generated
from the coupling of photons and baryons before recombination is
accounted for by Thomson scattering terms added to their respective
equations of motion in \S\S 5.5--5.7.  Even in the case of isocurvature
baryon models, which may have large perturbations in the entropy per
baryon ab initio, the corrections to $\delta P/\delta\rho=w$ are
generally very small because of the large photon to baryon ratio.

\section{Evolution Equations for Matter and Radiation}
\label{sec:boltz}
\subsection{Phase Space and the Boltzmann Equation}
A phase space is described by six variables: three positions $x^i$ and
their conjugate momenta $P_i$.  Our treatment of phase space is based
on the time-slicing of a definite gauge (synchronous or conformal
Newtonian).  Although this approach is not manifestly covariant, it
yields correct results provided the gauge-dependent quantities are converted
to observables at the end of the computation.

The conjugate momentum has the property that it is simply the spatial
part of the 4-momentum with lower indices, i.e., for a particle of mass
$m$, $P_i=mU_i\,,$ where $U_i=dx_i/\sqrt{-ds^2}$.  One can verify that
the conjugate momentum is related to the proper momentum $p^i=p_i$ measured
by an observer at a fixed spatial coordinate value by
\begin{eqnarray}
\label{psubi}
 	P_i &=a(\delta_{ij}+\frac{1}{2}h_{ij}) p^j\,,\qquad
	 &\mbox{in synchronous gauge}\,, \nonumber\\
	P_i &=a(1-\phi)p_i\,,\qquad
	 &\mbox{in conformal Newtonian gauge}\,.
\end{eqnarray}
In the absence of metric perturbations, Hamilton's
equations imply that the conjugate momenta are constant, so the proper
momenta redshift as $a^{-1}$.

The phase space distribution of the particles gives the
number of particles in a differential volume $dx^1 dx^2 dx^3 dP_1 dP_2
dP_3$ in phase space:
\begin{equation}
	f(x^i,P_j,\tau)dx^1 dx^2 dx^3 dP_1 dP_2 dP_3 = dN \ .
\end{equation}
Importantly, $f$ is a scalar and is invariant under canonical
transformations.  The zeroth-order phase space distribution is the
Fermi-Dirac distribution for fermions ($+$ sign) and the Bose-Einstein
distribution for bosons ($-$ sign):
\begin{equation}
\label{equil-dist}
   f_0=f_0(\epsilon) = {g_s\over h_{\rm P}^3}{1\over e^{\epsilon/
     k_{\rm B} T_0} \pm 1}\,,
\end{equation}
where $\epsilon = a(p^2+m^2)^{1/2} =  (P^2 + a^2m^2)^{1/2}\,$, $T_0 = a T$
denotes the temperature of the particles today, the factor $g_s$ is the
number of spin degrees of freedom, and $h_{\rm P}$ and $k_{\rm B}$ are
the Planck and the Boltzmann constants.

When the spacetime is perturbed, $x^i$ and $P_i$ remain canonically
conjugate variables, with equations of motion given by Hamilton's
equations (Bertschinger 1993).  However, following common practice
(e.g., Bond \& Szalay 1983) we shall find it convenient to replace
$P_j$ by $q_j\equiv ap_j$ in order to eliminate the metric perturbations
from the definition of the momenta.  Moreover, we shall write the
comoving 3-momentum $q_j$ in terms of its magnitude and direction:
$q_j=qn_j$ where $n^in_i=\delta_{ij}n^in^j=1$.  Thus, we change our
phase space variables, replacing $f(x^i,P_j,\tau)$ by $f(x^i,q,n_j,\tau)$.
While this is not a canonical transformation (i.e., $q_i$ is not the
momentum conjugate to $x^i$), it is perfectly valid provided that we
correctly transform the momenta in Hamilton's equations.  Note that we
do not transform $f$.  Because $q_j$ are not the conjugate momenta,
$d^3xd^3q$ is not the phase space volume element, and $fd^3xd^3q$ is
not the particle number.  In the conformal Newtonian gauge, for example,
$(1-3\phi)fd^3xd^3q$ is the particle number; this result is sensible
because $a(1-\phi)dx^i$ is the proper distance.

In the perturbed case we shall continue to define $\epsilon$ as $a(\tau)$
times the proper energy measured by a comoving observer,  $\epsilon=
(q^2+a^2m^2)^{1/2}$.  This is related to the time component of the
4-momentum by $P_0=-\epsilon$ in the synchronous gauge and $P_0=-(1+\psi)
\epsilon$ in the conformal Newtonian gauge.  For the CDM+HDM models we
are interested in, the photons, the massless neutrinos, and the massive
neutrinos at the time of neutrino decoupling are all ultra-relativistic
particles, so $\epsilon$ in the unperturbed Fermi-Dirac and Bose-Einstein
distributions can be simply replaced by the new variable $q$.

The general expression for the energy-momentum tensor written
in terms of the distribution function and the 4-momentum components
is given by
\begin{equation}
\label{tmunu}
	T_{\mu\nu}= \int dP_1 dP_2 dP_3\,(-g)^{-1/2}\,
	{P_\mu P_\nu\over P^0} f(x^i,P_j,\tau) \,,
\end{equation}
where $g$ denotes the determinant of $g_{\mu\nu}$.
It is convenient to write the phase space distribution as a
zeroth-order distribution plus a perturbed piece in
the new variables $q$ and $n_j$:
\begin{equation}
\label{f-pert}
	f(x^i,P_j,\tau) = f_0(q) \left[ 1 + \Psi(x^i,q,n_j,\tau) \right] \,.
\end{equation}

In the synchronous gauge, we have $(-g)^{-1/2} = a^{-4}(1-\frac{1}{2}h)$
and $dP_1 dP_2 dP_3 = (1+\frac{1}{2}h) q^2 dq d\Omega$
to linear order, where $h \equiv h_{ii}$ and $d\Omega$ is the solid
angle associated with direction $n_i$.
Using the relations $\int d\Omega\,n_i n_j = 4\pi \delta_{ij}/3$ and
$\int d\Omega\,n_i = \int d\Omega\,n_i n_j n_k = 0$, it then follows
from equation (\ref{tmunu}) that
\begin{eqnarray}
\label{tmunu2}
	T^0{}_{\!0} &=& - a^{-4} \int q^2dq\,d\Omega\,
		\sqrt{q^2+m^2a^2}\,f_0(q)\,(1+\Psi) \,,\nonumber\\
	T^0{}_{\!i} &=& a^{-4} \int q^2dq\,d\Omega\,
		q\,n_i\,f_0(q)\,\Psi \,,\\
	T^i{}_{\!j} &=& a^{-4} \int q^2dqd\Omega
		\,{q^2 n_i n_j\over \sqrt{q^2+m^2a^2}}\,f_0(q)\,(1+\Psi)
		\nonumber
\end{eqnarray}
to linear order in the perturbations.  Note that we have eliminated the
explicit dependence on the metric perturbations in equation (\ref{tmunu})
by redefining $P_i$ in terms of $q$ and $n_i$.   Note also that the
comoving energy $\epsilon(q,\tau)=(q^2+a^2m^2)^{1/2}$ is used in the
integrands but not in the argument of the unperturbed distribution function.

In the conformal Newtonian gauge, $(-g)^{-1/2} = a^{-4}(1-\psi+3\phi)$
and $dP_1 dP_2 dP_3 = (1-3\phi) q^2 dq d\Omega$.  It then follows that
the components of the energy-momentum tensor have the
same form as in equations (\ref{tmunu2}).  Of course, it is understood that
the variables $q$ and $n_i$ in this case are defined in relation to
the conjugate momentum $P_i$ in the conformal Newtonian coordinates,
and not the synchronous coordinates.  (They differ because comoving
observers in the two coordinate systems are not the same.) The expansion
factor $a$ and $\Psi$ are evaluated at the coordinates $(x^i,\tau)$
in the conformal Newtonian gauge.

The phase space distribution evolves according to the Boltzmann equation.
In terms of our variables $(x^i,q,n_j,\tau)$ this is
\begin{equation}
	{Df \over d\tau} = {\partial f \over \partial \tau}
	+ {dx^i \over d\tau}{\partial f\over \partial x^i}
	+ {dq \over d\tau}{\partial f\over \partial q}
	+ {dn_i \over d\tau}{\partial f\over \partial n_i}
	= \left( {\partial f \over \partial\tau} \right)_C\,,
\end{equation}
where the right-hand side involves terms due to collisions,
whose form depends on the type of particle interactions involved.
{}From the geodesic equation
\begin{equation}
	P^0 {dP^\mu \over d\tau} + \Gamma^\mu{}_{\!\alpha\beta}
	\,P^\alpha P^\beta = 0 \,,
\end{equation}
it is straightforward to show that
\begin{eqnarray}
	dq/d\tau &=& -{1\over 2} q\dot{h}_{ij}n_i n_j \qquad \mbox{in
	synchronous gauge}\,, \nonumber\\
	dq/d\tau &=& q\dot{\phi}-\epsilon(q,\tau)\,n_i \partial_i \psi \qquad
	\mbox{in conformal Newtonian gauge}\,,
\end{eqnarray}
and $dn_i/d\tau$ is $O(h)$.  Since $\partial f/\partial n_i$
is also a first-order quantity, the term $(d n_i/d\tau)(\partial
f/\partial n_i)$ in the Boltzmann equation can be neglected to first
order.  Then the Boltzmann equation in $k$-space can be written as
\medskip
\newline {\it Synchronous gauge ---\hfil}
\begin{equation}
\label{bolt-syn}
  {\partial \Psi \over \partial \tau} + i\,{q\over\epsilon}
	(\vec{k}\cdot \hat{n})\Psi +
     {d\ln f_0 \over d\ln q}\, \left[\dot{\eta} - {\dot{h}+6\dot{\eta}
       \over 2}(\hat{k}\cdot\hat{n})^2 \right] =
     {1\over f_0}\,\left( {\partial f \over \partial\tau} \right)_C \,,
\end{equation}
\newline {\it Conformal Newtonian gauge ---\hfil}
\begin{equation}
\label{bolt-con}
  {\partial \Psi \over \partial \tau} + i\,{q \over
	\epsilon}\,(\vec{k}\cdot \hat{n})\,\Psi +
     {d\ln f_0 \over d\ln q} \left[\dot{\phi} - i\,{\epsilon\over q}
       (\vec{k}\cdot\hat{n})\,\psi \right] =
   {1\over f_0}\,\left( {\partial f \over \partial\tau} \right)_C \,.
\end{equation}
Equations (\ref{bolt-syn}) and (\ref{bolt-con}) can also be derived using
the canonical phase space variables $x^i$ and $P_j$ and Hamilton's equations
(instead of the geodesic equation), followed by a transformation from $P_j$
to $qn_j$.

The terms in the Boltzmann equation depend on the direction of the
momentum $\hat{n}$ only through its angle with $\vec{k}$.  (We shall see
that this is true of the collision term for photons as well as
the convective and metric perturbation terms.) Therefore, if the
momentum-dependence of the initial phase space perturbation is axially
symmetric about $\vec k$, it will remain axially symmetric.
If axially-asymmetric perturbations in the neutrinos or other
collisionless particles are produced, they would generate no scalar
metric perturbations and thus would have no effect on other species.
Therefore, we shall assume that the initial momentum-dependence is
axially symmetric so that $\Psi$ depends on $\vec{q}=q\hat{n}$ only
through $q$ and $\hat{k}\cdot\hat{n}$.  This assumption, which
effectively reduces the dimensionality of phase space perturbations by
one (after Fourier transforming on the spatial coordinates), has been
made (implicitly, if not explicitly) in all previous studies of the
evolution of scalar perturbations.

\subsection{Cold Dark Matter}
CDM interacts with other particles only through gravity and can be
treated as a pressureless perfect fluid.  The CDM particles can be
used to define the synchronous coordinates and therefore have zero
peculiar velocities in this gauge.  Setting $\theta=\sigma=0$ and
$w=\dot{w}=0$ in equations (\ref{fluid}) leads to
\medskip
\newline {\it Synchronous gauge ---\hfil}
\begin{equation}
\label{cdm}
	\dot{\delta_c} = -\frac{1}{2}\,\dot{h} \,.
\end{equation}
The CDM fluid velocity in the conformal Newtonian gauge, however, is
not zero in general.  In $k$-space, equations (\ref{fluid2}) give
\medskip
\newline {\it Conformal Newtonian gauge ---\hfil}
\begin{equation}
\label{cdm2}
	\dot{\delta_c} = -\theta_c + 3\dot{\phi}\,, \quad
	\dot{\theta}_c = - {\dot{a}\over a}\,\theta_c+k^2\psi \,.
\end{equation}
The subscript $c$ in $\delta_c$ and $\theta_c$ denotes the cold dark matter.

\subsection{Massless Neutrinos}
\label{sec:lessnu}
The energy density and the pressure for massless neutrinos
(labeled by subscripts $\nu$) are $\rho_\nu = 3P_\nu = -T^0{}_{\!0} =
T^i{}_{\!i}$.  From equations (\ref{tmunu2}) the unperturbed
energy density $\bar{\rho}_\nu$ and pressure $\bar{P}_\nu$ are given by
\begin{equation}
	 \bar{\rho}_\nu = 3 \bar{P}_\nu = a^{-4}
	\int q^2dq d\Omega\,q f_0(q) \,,
\end{equation}
and the perturbations of energy density $\delta\rho_\nu$, pressure $\delta
P_\nu$, energy flux $\delta T^0_{\nu\,i}\,$, and shear stress
$\Sigma^i_{\nu\,j}=T^i_{\nu\,j}-P_\nu\delta_{ij}$ are given by
\begin{eqnarray}
     \delta\rho_\nu &=& 3 \delta P_\nu = a^{-4}
        \int q^2dq d\Omega\,q f_0(q) \Psi \,,\nonumber\\
     \delta T^0_{\nu\,i} &=& a^{-4}
	\int q^2dq d\Omega\,qn_i\,f_0(q) \Psi \,,\\
     \Sigma^i_{\nu\,j} &=& a^{-4}
	\int q^2dq d\Omega\,q(n_in_j-\frac{1}{3}\delta_{ij})\,f_0(q)\Psi \,.
	\nonumber
\end{eqnarray}
The unperturbed energy flux and shear stress are zero.

The Boltzmann equation simplifies for massless particles, for which
$\epsilon=q$.  To reduce the number of variables we integrate out the
$q$-dependence in the neutrino distribution function and expand the
angular dependence of the perturbation in a series of Legendre polynomials
$P_l(\hat{k}\cdot\hat{n})$:
\begin{equation}
\label{fsubl}
      F_\nu(\vec{k},\hat{n},\tau) \equiv {\int q^2 dq\,q f_0(q)\Psi
	\over \int q^2 dq\,q f_0(q)} \equiv \sum_{l=0}^\infty(-i)^l
	(2l+1)F_{\nu\,l}(\vec{k},\tau)P_l(\hat{k}\cdot\hat{n})\,.
\end{equation}
As noted in \S 5.1, the dependence on $\hat n$ arises only through
$\hat{k}\cdot\hat{n}$, so that a general distribution may be represented
as in equation (\ref{fsubl}).  The factor $(-i)^l(2l+1)$ is chosen to
simplify the expansion of a plane wave: $F_\nu=\exp(-i\vec k\cdot\vec
x\,)$ with $\vec x=r(\tau)\hat n$ has expansion coefficients
$F_{\nu\,l}=j_l(kr)$ given by spherical Bessel functions.

In terms of the new variable $F_\nu(\vec k,\hat n,\tau)$ and its harmonic
expansion coefficients, the perturbations $\delta_\nu$,
$\bar{\rho}_\nu$, $\theta_\nu$, and $\sigma_\nu$  (defined in eq.
[\ref{theta}]) take the form
\begin{eqnarray}
\label{fsubnu}
      \delta_\nu &=& {1\over 4\pi} \int d\Omega
		F_\nu(\vec{k},\hat{n},\tau)=F_{\nu\,0}\,,\nonumber\\
      \theta_\nu &=& {3i \over 16\pi} \int d\Omega\,
		(\vec{k}\cdot\hat{n}) F_\nu(\vec{k},\hat{n},\tau)
		={3\over 4} k F_{\nu\,1}\,,\\
      \sigma_\nu &=& -{3 \over 16\pi} \int d\Omega
	\left[(\hat{k}\cdot\hat{n})^2 - {1\over 3}\right]
	 F_\nu(\vec{k},\hat{n},\tau)={1\over 2} F_{\nu\,2}\,,\nonumber
\end{eqnarray}
where we have used $\rho_\nu=3P_\nu$ for the massless neutrinos.

Integrating equations (\ref{bolt-syn}) and (\ref{bolt-con}) over
$q^2dq\,q f_0(q)$ and dividing them by $\int q^2dq\,q f_0(q)$, the
Boltzmann equation for massless neutrinos becomes
\begin{eqnarray}
\label{bolmn}
  {\partial F_\nu\over\partial\tau}+ik\mu F_\nu &=&
    -\frac{2}{3}\dot h-\frac{4}{3}(\dot h+6\dot\eta)P_2(\mu)\qquad
      \mbox{in synchronous gauge}\,, \nonumber\\
  {\partial F_\nu\over\partial\tau}+ik\mu F_\nu &=&
    4\,(\dot\phi-ik\mu\psi)\qquad
      \mbox{in conformal Newtonian gauge}\,,
\end{eqnarray}
where $\mu\equiv\hat k\cdot\hat n$ and $P_2(\mu)=\frac{1}{2}(3\mu^2-1)$
is the Legendre polynomial of degree 2.  Substituting the Legendre
expansion for $F_\nu$ and using the orthonormality of the Legendre
polynomials and the recursion relation $(l+1)P_{l+1}(\mu) = (2l+1)
\mu P_l(\mu) -l P_{l-1}(\mu)$, we obtain
\medskip
\newline {\it Synchronous gauge ---\hfil}
\begin{eqnarray}
\label{massless}
	\dot{\delta}_\nu &=& -{4\over 3}\theta_\nu
		-{2\over 3}\dot{h} \,,\nonumber\\
	\dot{\theta}_\nu &=& k^2 \left(\frac{1}{4}\delta_\nu
		- \sigma_\nu \right)\,,\nonumber\\
	\dot{F}_{\nu\,2} &=& 2\dot\sigma_\nu = {8\over15}\theta_\nu
	    	- {3\over 5} k F_{\nu\,3} + {4\over15}\dot{h}
	  	+ {8\over5} \dot{\eta} \,,\nonumber\\
	\dot{F}_{\nu\,l} &=& {k\over2l+1}\left[ l
	 	F_{\nu\,(l-1)} - (l+1)
		F_{\nu\,(l+1)} \right]\,, \quad l \geq 3 \,.
\end{eqnarray}
\newline {\it Conformal Newtonian gauge ---\hfil}
\begin{eqnarray}
\label{massless2}
	\dot{\delta}_\nu &=& -{4\over 3}\theta_\nu
		+4\dot{\phi} \,, \nonumber\\
	\dot{\theta}_\nu &=&k^2 \left(\frac{1}{4}\delta_\nu
		- \sigma_\nu \right)
		+ k^2\psi \,, \nonumber\\
	\dot{F}_{\nu\,l} &=& {k\over2l+1}\left[ l
 		F_{\nu\,(l-1)} - (l+1)
		F_{\nu\,(l+1)} \right]\,, \quad l \geq 2 \,.
\end{eqnarray}
This set of equations governs the evolution of the phase space
distribution of massless neutrinos.  Note that a given mode $F_l$ is
coupled only to the $(l-1)$ and $(l+1)$ neighboring modes.

The Boltzmann equation (\ref{bolmn}) has been transformed into an
infinite hierarchy of moment equations that must be truncated at
some maximum multipole order $l_{\rm max}$.  One simple but inaccurate
method is to set $F_{\nu\,l}=0$ for $l>l_{\rm max}$.  The problem with
this scheme is that the coupling of multipoles in equations
(\ref{massless}) and (\ref{massless2}) leads to the propagation of
errors from $l_{\rm max}$ to smaller $l$.  Indeed, these errors can
propagate to $l=0$ in a time $\tau\approx l_{\rm max}/k$ and then
reflect back to increasing $l$, leading to amplification of errors in
the ``square well'' $0\le l\le l_{\rm max}$.

An improved truncation scheme is based on extrapolating the behavior
of $F_{\nu\,l}$ to $l=l_{\rm max}+1$.  For $l>1$, the multipole
moments are gauge-invariant and numerical solutions show that they
exhibit damped oscillations reminiscent of spherical Bessel functions.
In fact, if $\partial_\tau(\phi+\psi)=0$ (choosing conformal Newtonian
gauge for simplicity), the time dependence of the exact solution of the
Boltzmann hierarchy is $F_{\nu\,l}(k,\tau)\propto j_l(k\tau)$ for $l>0$.
If we assume that this relation holds approximately for time-varying
potentials, using a recurrence relation for spherical Bessel functions
we get
\begin{equation}
\label{truncnu}
     F_{\nu\,(l_{\rm max}+1)}\approx{(2l_{\rm max}+1)\over k\tau}\,F_{\nu
       \,l_{\rm max}}-F_{\nu\,(l_{\rm max}-1)}\ .
\end{equation}
By numerically integrating the Boltzmann hierarchy with different
choices of $l_{\rm max}$, we have found that this truncation scheme
greatly improves on the simple truncation $F_{\nu\,(l_{\rm max}+1)}=0$.
However, time-variations of the potentials during the
radiation-dominated era make even equation (\ref{truncnu}) a poor
approximation if $l_{\max}$ is chosen too small.  We shall discuss
in \S 8 our choices for $l_{\rm max}$.

\subsection{Massive Neutrinos}

Massive neutrinos also obey the collisionless Boltzmann equation.
The evolution of the distribution function for massive neutrinos
is, however, complicated by their nonzero mass.  From equations
(\ref{tmunu2}), the unperturbed energy density and pressure for
massive neutrinos (labeled by subscripts ``$h$'' for HDM) are given by
\begin{equation}
	\bar{\rho}_h = a^{-4} \int q^2dq\,d\Omega\,
		\epsilon f_0(q) \,,\qquad
	\bar{P}_h = {1\over 3} a^{-4} \int q^2dq\,d\Omega\,
		{q^2\over\epsilon} f_0(q) \,,
\end{equation}
where $\epsilon = \epsilon(q,\tau) = \sqrt{q^2 + m_\nu^2a^2}$, while
the perturbations are
\begin{eqnarray}
\label{delta}
   \delta\rho_h =& a^{-4} \int q^2dq\,d\Omega\,\epsilon f_0(q) \Psi
	\,,\qquad  \delta P_h =& {1\over 3} a^{-4}
        \int q^2dq\,d\Omega\,{q^2\over \epsilon} f_0(q) \Psi \,,\nonumber\\
     \delta T^0_{h\,i} =& a^{-4}
	\int q^2dq\,d\Omega\,q n_i\,f_0(q) \Psi \,,\qquad
     \Sigma^i_{h\,j} =& a^{-4}
	\int q^2dq\,d\Omega\,{q^2\over\epsilon}
	(n_in_j-{1\over 3}\delta_{ij})\,f_0(q) \Psi \,.
\end{eqnarray}
Since the comoving energy-momentum relation $\epsilon(q,\tau)$ depends
on both the momentum and time, we can not simplify the calculations by
integrating out the $q$-dependence in the distribution function as we
did for the massless neutrinos above (see eq. [\ref{fsubl}]).  Instead of
applying equation (\ref{fsubl}), we expand the perturbation $\Psi$
directly in a Legendre series
\begin{equation}
\label{psisubl}
      \Psi(\vec{k},\hat{n},q,\tau)
	= \sum_{l=0}^\infty (-i)^l(2l+1) \Psi_l(\vec{k},q,\tau)
	P_l(\hat{k}\cdot\hat{n})\,.
\end{equation}
Then the perturbed energy density, pressure, energy flux, and shear
stress in $k$-space are given by
\begin{eqnarray}
\label{delta2}
   \delta\rho_h &=& 4\pi a^{-4}
        \int q^2 dq\,\epsilon f_0(q) \Psi_0 \,, \nonumber\\
   \delta P_h &=& {4\pi \over 3} a^{-4}
        \int q^2 dq\,{q^2\over \epsilon} f_0(q) \Psi_0
	 \,, \nonumber\\
   (\bar\rho_h +\bar P_h) \theta_h &=& 4\pi k a^{-4}
	\int q^2 dq\,q f_0(q)\Psi_1 \,, \nonumber\\
   (\bar\rho_h +\bar P_h) \sigma_h &=& {8\pi\over 3} a^{-4}
	\int q^2 dq\,{q^2\over\epsilon} f_0(q) \Psi_2 \,.
\end{eqnarray}

Following the same procedure used for the massless neutrinos, the Boltzmann
equation becomes
\medskip
\newline{\it Synchronous gauge ---\hfil}
\begin{eqnarray}
\label{massive}
     \dot{\Psi}_0 &=& -{qk\over \epsilon}\Psi_1
  	  +{1\over 6}\dot{h} {d\ln f_0\over d\ln q}
		\,, \nonumber\\
     \dot{\Psi}_1 &=& {qk\over 3\epsilon} \left(\Psi_0
		      - 2 \Psi_2 \right) \,, \nonumber\\
     \dot{\Psi}_2 &=& {qk\over 5\epsilon} \left(
	2\Psi_1 - 3\Psi_3 \right)
	 - \left( {1\over15}\dot{h} + {2\over5} \dot{\eta} \right)
	{d\ln f_0\over d\ln q} \,,\\
    \dot{\Psi}_l &=& {qk \over (2l+1)\epsilon} \left[ l\Psi_{l-1}
        - (l+1)\Psi_{l+1} \right]\,,
	\quad l \geq 3 \,. \nonumber
\end{eqnarray}
\newline {\it Conformal Newtonian gauge ---\hfil}
\begin{eqnarray}
\label{massive2}
     \dot{\Psi}_0 &=& -{qk\over\epsilon}\Psi_1
  	  -\dot{\phi} {d\ln f_0\over d\ln q} \,, \nonumber\\
     \dot{\Psi}_1 &=& {qk\over 3\epsilon} \left(\Psi_0
	  - 2 \Psi_2 \right)
	  - {\epsilon\,k\over 3q} \psi {d\ln f_0\over d\ln q} \,,\\
    \dot{\Psi}_l &=& {qk \over (2l+1)\epsilon} \left[
	l\Psi_{l-1} - (l+1)\Psi_{l+1} \right]\,,
	\quad l \geq 2 \,. \nonumber
\end{eqnarray}
Because of the $q$-dependence in these equations, it requires much
more computing time to carry out the time integration for the massive
neutrino.  Bond \& Szalay (1983) used a 16-point Gauss-Legendre method
to approximate the $q$-integration.  We do not use this method and
instead perform the integration using sixth- or eighth-order Newton-Cotes
quadrature (plus a remainder obtained by asymptotic expansion) with a
$q$-grid of 128 $q$-points for every wavenumber $k$.  We verified that
this was enough to ensure a relative accuracy no worse than $10^{-4}$
by trying the integration with twice as many points.  Then the
perturbations $\delta\rho_h$, $\delta P_h$, $\theta_h$, and $\sigma_h$
that enter the right-hand side of the Einstein equations are calculated
from equations (\ref{delta2}) by numerically integrating $\Psi_0,
\Psi_1$ and $\Psi_2$ over $q$.

As with massless neutrinos, we must truncate the Boltzmann hierarchy
for massive neutrinos.  We have found the following scheme to work well:
\begin{equation}
\label{truncmnu}
     \Psi_{\nu\,(l_{\rm max}+1)}\approx{(2l_{\rm max}+1)\epsilon\over
       qk\tau}\,\Psi_{\nu\,l_{\rm max}}-\Psi_{\nu\,(l_{\rm max}-1)}\ .
\end{equation}
Because the higher multipole moments decay rapidly once the neutrinos
become nonrelativistic, it is possible to choose a much smaller $l_{\rm
max}$ for massive neutrinos than for massless ones.

\subsection{Photons}
Photons evolve differently before and after recombination.  Before
recombination, photons and baryons are tightly coupled, interacting
mainly via Thomson scattering (and the electrostatic coupling of electrons
and ions).  In Thomson scatterings, the photon energy $h\nu$ is assumed
to be much less than the electron rest mass $m_e \sim 0.511$ MeV and
the recoil of the electron in the initial electron rest frame is
neglected.  (We are concerned with the period after neutrino decoupling,
when $T < m_e$.) The classical differential cross section for Thomson
scattering is given by $d\sigma/d\Omega = 3\sigma_T(1+\cos^2\theta)
/16\pi$, where $\sigma_T = 0.6652 \times 10^{-24}$ cm$^2$ and $\theta$
is the scattering angle (e.g., Jackson 1975).  After recombination,
the universe gradually becomes transparent to radiation and photons
travel almost freely, although Thomson scattering continues to
transfer energy and momentum between the photons and the matter.

The evolution of the photon distribution function can be treated
in a similar way as the massless neutrinos, with the exception that
the collisional terms on the right-hand side of the Boltzmann equation
are now present and they depend on polarization.  Photons propagating
in direction $\hat{n}$ are linearly polarized in the plane perpendicular
to $\hat{n}$ due to scattering of electron density perturbations with
wavevector $\vec k$.  We shall track both the sum (total intensity) and
difference (Stokes parameter $Q$) of the phase space densities in the
two polarization states for each $\vec k$ and $\hat{n}$.  We shall denote
the former (the momentum-averaged total phase space density perturbation,
summed over polarizations) by $F_\gamma(\vec k,\hat{n},\tau)$, defined as
in equation (\ref{fsubl}).  Similarly, $G_\gamma(\vec k,\hat{n},\tau)$
is the difference of the two linear polarization components.  The linearized
collision operators for Thomson scattering are (Bond \& Efstathiou 1984,
1987; Kosowsky 1995)
\begin{equation}
\label{scatop}
	\left( {\partial F_\gamma \over \partial\tau} \right)_C
	    = a n_e \sigma_T \left[ -F_\gamma
	    + F_{\gamma\,0} + 4\hat{n}\cdot\vec{v}_e
	    - {1\over 2}\left(F_{\gamma\,2}+G_{\gamma\,0}+G_{\gamma\,2}\right)
                  P_2     \right] \,,
\end{equation}
\begin{equation}
	\left( {\partial G_\gamma \over \partial\tau} \right)_C
	    = a n_e \sigma_T \left[ -G_\gamma
	    + {1\over 2}\left(F_{\gamma\,2}+G_{\gamma\,0}+G_{\gamma\,2}\right)
                  \left(1-P_2\right)
     \right] \,,
\end{equation}
where $n_e$ and $\vec{v}_e$ are the proper mean density and
velocity of the electrons.  The terms proportional to $P_2$ come from
the polarization-dependence of the Thomson cross section which, when
averaged over incident directions, give an angular dependence
$1+\cos^2\theta$ in the Thomson cross section even for unpolarized
radiation.  The anisotropic scattering and net polarization were both
neglected by Peebles \& Yu (1970); the former (but not the latter) were
included by Wilson \& Silk (1980, 1981) and most later workers.

Expanding $F_\gamma (\vec{k},\hat{n},\tau)$ and $G_\gamma(\vec{k},
\hat{n},\tau)$ in Legendre series as in equation (\ref{fsubl}) and
using the relations $\hat{n}\cdot\vec{v}_e = -(i\theta_b/k)P_1(\hat{k}
\cdot\hat{n})$, $F_{\gamma\,1}=4\theta_\gamma/(3k)$, and $F_{\gamma\,2}=
2\sigma_\gamma$, the collision operators can be rewritten as
\begin{equation}
\label{thom1}
	\left( {\partial F_\gamma \over \partial\tau} \right)_C
	    = a n_e \sigma_T \left[ {4i\over k}
	    (\theta_\gamma-\theta_b)P_1 + \left(9\sigma_\gamma - {1\over 2}
          G_{\gamma\,0} - {1\over 2} G_{\gamma\,2}\right) P_2
	    - \sum_{l\ge 3}^\infty (-i)^l(2l+1) F_{\gamma\,l} P_l \right]\,,
\end{equation}
\begin{equation}
\label{thom2}
	\left( {\partial G_\gamma \over \partial\tau} \right)_C
	    = a n_e \sigma_T \left[ {1\over 2} \left(
	    F_{\gamma\,2} + G_{\gamma\,0}+ G_{\gamma\,2}\right)(1-P_2)
	    - \sum_{l\ge 0 }^\infty (-i)^l(2l+1) G_{\gamma\,l} P_l \right]\,.
\end{equation}
The left-hand-side of the Boltzmann equation for $F_{\gamma}$ and $G_{\gamma}$
remain the same as for the massless neutrinos, so we obtain
\medskip
\newline {\it Synchronous gauge ---\hfil}
\begin{eqnarray}
\label{photon}
     \dot{\delta}_\gamma &=& -{4\over 3}\theta_\gamma
	-{2\over 3}\dot{h} \,,\nonumber\\
     \dot{\theta}_\gamma &=& k^2 \left(\frac{1}{4}\delta_\gamma
	- \sigma_\gamma\right)
	+ a n_e \sigma_T (\theta_b - \theta_\gamma) \,,\nonumber\\
     \dot{F}_{\gamma\,2} &=& 2\dot\sigma_\gamma={8\over15}\theta_\gamma
       - {3\over 5} k F_{\gamma\,3} + {4\over15}\dot{h}
       + {8\over5} \dot{\eta} - {9\over5} an_e \sigma_T \sigma_\gamma
       + {1\over10} an_e \sigma_T \left(G_{\gamma\,0}+G_{\gamma\,2}\right)
	\,, \nonumber\\
     \dot{F}_{\gamma\,l} &=& {k\over2l+1}\left[ l
	F_{\gamma\,(l-1)} - (l+1) F_{\gamma\,(l+1)}\right]
	- an_e \sigma_T F_{\gamma\,l} \,,\quad l \geq 3 \,,\nonumber\\
     \dot{G}_{\gamma\,l} &=& {k\over2l+1}\left[ l
        G_{\gamma\,(l-1)} - (l+1) G_{\gamma\,(l+1)}\right]
        + an_e \sigma_T \left[ -G_{\gamma\,l} + {1\over2} \left(
	F_{\gamma\,2}+G_{\gamma\,0}+G_{\gamma\,2}\right)\left(
	\delta_{l0}+{\delta_{l2}\over5}\right)\right] \,,\nonumber\\
\end{eqnarray}
\newline {\it Conformal Newtonian gauge ---\hfil}
\begin{eqnarray}
\label{photon2}
     \dot{\delta}_\gamma &=& -{4\over 3}\theta_\gamma
	+4\dot{\phi} \,,\nonumber\\
     \dot{\theta}_\gamma &=& k^2 \left(\frac{1}{4}\delta_\gamma
	- \sigma_\gamma \right) + k^2 \psi
	+ a n_e \sigma_T (\theta_b-\theta_\gamma) \,,\nonumber\\
     \dot{F}_{\gamma\,2} &=& 2\dot\sigma_\gamma= {8\over15}\theta_\gamma
        - {3\over 5} k F_{\gamma\,3}
	- {9\over5} an_e \sigma_T \sigma_\gamma
       + {1\over10} an_e \sigma_T \left(G_{\gamma\,0}+G_{\gamma\,2}\right)
 \,, \nonumber\\
     \dot{F}_{\gamma\,l} &=& {k\over2l+1}\left[ l
	F_{\gamma\,(l-1)} - (l+1)F_{\gamma\,(l+1)}
	\right] - an_e \sigma_T F_{\gamma\,l} \,,\quad l \geq 3 \,\nonumber\\
     \dot{G}_{\gamma\,l} &=& {k\over2l+1}\left[ l
        G_{\gamma\,(l-1)} - (l+1) G_{\gamma\,(l+1)}\right]
        + an_e \sigma_T \left[ -G_{\gamma\,l} + {1\over2} \left(
	F_{\gamma\,2}+G_{\gamma\,0}+G_{\gamma\,2}\right)\left(
	\delta_{l0}+{\delta_{l2}\over5}\right)\right] \,,\nonumber\\
\end{eqnarray}
The subscripts $\gamma$ and $b$ refer to photons and baryons
respectively.

We truncate the photon Boltzmann equations in a manner similar to
massless neutrinos (eq.~\ref{truncnu}), except that Thomson opacity terms
must be added.  For $l=l_{\rm max}$ we replace equations (\ref{photon})
and (\ref{photon2}) by
\begin{eqnarray}
\label{truncphot}
     \dot{F}_{\gamma\,l} &=& k F_{\gamma\,(l-1)} - {l+1\over\tau}
        F_{\gamma\,l} - an_e \sigma_T F_{\gamma\,l} \,,\nonumber\\
     \dot{G}_{\gamma\,l} &=& k G_{\gamma\,(l-1)} - {l+1\over\tau}
        G_{\gamma\,l} - an_e \sigma_T G_{\gamma\,l} \,.
\end{eqnarray}

\subsection{Baryons}
The baryons (and electrons) behave like a non-relativistic fluid described,
in the absence of coupling to radiation, by the energy-momentum conservation
equations (\ref{fluid}) and (\ref{fluid2}) with $\delta P_b/\delta\rho_b=
c_s^2=w\ll1$ and $\sigma=0$.  Since the baryons are very nonrelativistic
after neutrino decoupling (the period of interest), we may neglect $w$
and $\delta P/\delta\rho$ in all terms except the acoustic term $c_s^2k^2
\delta$ (which is important for sufficiently high $k$; note that the shear
stress term $k^2\sigma$ is far smaller so we neglect it).  Before
recombination, however, the coupling of the baryons and the photons
causes a transfer of momentum and energy between the two components.

{}From equation (\ref{theta}) the momentum density $T^0{}_{\!j}$ for a given
species is related to $\theta\,$ by $ik^j \delta T^0{}_{\!j} =
(\bar{\rho}+\bar{P})\theta$.  The momentum transfer into the
photon component is represented by $an_e\sigma_T(\theta_b -
\theta_\gamma)$ of equations (\ref{photon}) and (\ref{photon2}).
Momentum conservation in Thomson scattering then implies that a term
$(4\bar\rho_\gamma/3
\bar\rho_b)\,an_e\sigma_T(\theta_\gamma - \theta_b)$ has to be added
to the equation for $\dot{\theta}_b\,$ (where we have used $\bar{P}_b
\ll \bar{\rho}_b$), so equations (\ref{fluid}) and (\ref{fluid2}) are
modified to become \medskip
\newline {\it Synchronous gauge ---\hfil}
\begin{eqnarray}
\label{baryon}
	\dot{\delta}_b &=& -\theta_b - {1\over 2}\dot{h} \,, \nonumber\\
	\dot{\theta}_b &=& -{\dot{a}\over a}\theta_b
	+ c_s^2 k^2\delta_b + {4\bar\rho_\gamma \over 3\bar\rho_b}
	 an_e\sigma_T (\theta_\gamma-\theta_b)\,,
\end{eqnarray}
\newline {\it Conformal Newtonian gauge ---\hfil}
\begin{eqnarray}
\label{baryon2}
	\dot{\delta}_b &=& -\theta_b + 3\dot{\phi} \,, \nonumber\\
	\dot{\theta}_b &=& -{\dot{a}\over a}\theta_b
	+ c_s^2 k^2\delta_b + {4\bar\rho_\gamma \over 3\bar\rho_b}
	 an_e\sigma_T (\theta_\gamma-\theta_b) + k^2\psi\,.
\end{eqnarray}
The square of the baryon sound speed is evaluated from
\begin{equation}
\label{soundsp}
        c_s^2={\dot P_b\over\dot\rho_b}={k_{\rm B}T_b\over\mu}
	  \left(1-{1\over3}{d\ln T_b\over d\ln a}\right)\,,
\end{equation}
where $\mu$ is the mean molecular weight (including free electrons
and all ions of H and He) and, in the second equality, we have
neglected the slow time variation of $\mu$.  (This approximation is
adequate because even during recombination, when $\dot\mu$ is largest,
the baryons contribute very little to the pressure of the photon-baryon
fluid.)  The baryon temperature evolves according to
\begin{equation}
\label{tbary}
        \dot T_b=-2{\dot{a}\over a}T_b+{8\over3}{\mu\over m_e}
	    {\bar\rho_\gamma\over\bar\rho_b}an_e\sigma_T
	    \left(T_\gamma-T_b\right)\,.
\end{equation}
We assume that electron-ion collisions are rapid enough for kinetic
equilibrium to hold with a common temperature $T_b$ for electrons and
all baryon species.  Equation (\ref{tbary}) follows from the first law
of thermodynamics, $dQ=(3/2)d(P_b/\rho_b)+P_bd(1/\rho_b)$, with specific
heating rate $\dot Q=4(\bar\rho_\gamma/\bar\rho_b)an_e\sigma_Tk_{\rm B}
(T_\gamma-T_b)$.

\subsection{Tight-Coupling Approximation}

Before recombination the Thomson opacity is so large that photons and
baryons are tightly coupled, with $an_e\sigma_T\equiv\tau_c^{-1}\gg
\dot a/a\sim\tau^{-1}$.  The large values of the Thomson drag terms in
equations (\ref{photon})--(\ref{baryon2}) for $\dot\theta_\gamma$ and
$\dot\theta_b$ make them numerically difficult to solve.
Therefore, in this limit we shall follow the method of Peebles \& Yu
(1970) to obtain an alternative form of the equations valid for
$\tau_c/\tau\ll1$ and $k\tau_c\ll1$.  The starting point is the exact
equation obtained by combining the second of equations (\ref{photon2})
and (\ref{baryon2}),
\begin{equation}
\label{momentum}
  (1+R)\dot\theta_b+{\dot a\over a}\theta_b-c_s^2 k^2\delta_b-
    k^2R\left(\frac{1}{4}\delta_\gamma-\sigma_\gamma\right)+
    R(\dot\theta_\gamma-\dot\theta_b)=(1+R)k^2\psi\ ,
\end{equation}
where the right-hand side is included only in the conformal Newtonian
gauge; in the synchronous gauge it is set to zero.  We have defined
$R\equiv(4/3)\bar\rho_\gamma/\bar\rho_b$.  We shall see that the terms
proportional to $(\dot\theta_\gamma-\dot\theta_b)$ and $\sigma_\gamma$
may be neglected to lowest order in
$\hbox{max}\,[k\tau_c,\tau_c/\tau]$, with the result that the baryons
and photons behave like a single coupled fluid with sound speed $c_{pb}$.
(This should be distinguished from $c_s$, which denotes the sound
speed for the baryons only and not that for the coupled photon-baryon
fluid.)  However, we require a more accurate approximation to account
for the slip between the photon and baryon fluids.

{}From the second of equations (\ref{photon2}), we have
\begin{equation}
\label{thetabdot}
  \theta_b-\theta_\gamma=\tau_c\left[\dot\theta_\gamma-k^2\left(
    \frac{1}{4}\delta_\gamma-\sigma_\gamma\right)-k^2\psi\right]
\end{equation}
in the conformal Newtonian gauge; in the synchronous gauge we simply set
$\psi=0$.  Writing $\dot\theta_\gamma$ as $\dot\theta_b+(\dot\theta_\gamma
-\dot\theta_b)$ and using equation (\ref{momentum}), we get
\begin{equation}
\label{slip}
  \theta_b-\theta_\gamma={\tau_c\over1+R}\left[-{\dot a\over a}
    \theta_b+k^2\left(c_s^2\delta_b-\frac{1}{4}\delta_\gamma+
    \sigma_\gamma\right)+\dot\theta_\gamma-\dot\theta_b\right]\ ,
\end{equation}
a result that is valid in both gauges.  From the third of equations
(\ref{photon}), we have
\begin{equation}
\label{f2-tca}
  \sigma_\gamma=\frac{\tau_c}{9}\left(\frac{8}{3}\theta_\gamma
    +\frac{4}{3}\dot h+8\dot\eta-10\dot\sigma_\gamma-3k
    F_{\gamma\,3}\right)
\end{equation}
in the synchronous gauge; in the conformal Newtonian gauge one sets
$\dot h=\dot\eta=0$.  We see that $\sigma_\gamma\sim\delta_\gamma
\times\hbox{max}\,[k\tau_c,\tau_c/\tau]$ (the case $k\tau_c$ corresponding
to acoustic oscillations with $\theta_\gamma\sim k\delta_\gamma$).
Higher moments of the photon distribution are smaller by additional
powers of $k\tau_c$ and we shall neglect them in the limit of tight
coupling considered here.  Our goal is to obtain equations for
$\dot\theta_b$ and $\dot\theta_\gamma$ that are accurate to second order
in $\tau_c$.

To get an equation for $\dot\theta_b$ we differentiate equation
(\ref{slip}) and use equation (\ref{momentum}).  Assuming that the gas
is nearly fully ionized so that $n_e\propto a^{-3}$ and that the baryon
temperature is approximately the radiation temperature implying $c_s^2
\propto a^{-1}$, we obtain
\begin{equation}
\label{sliprate}
  \dot\theta_b-\dot\theta_\gamma={2R\over1+R}{\dot a\over a}
    (\theta_b-\theta_\gamma)+{\tau_c\over1+R}\left[-{\ddot a\over a}
    \theta_b-{\dot a\over a}k^2\left(\frac{1}{2}\delta_\gamma+\psi\right)
    +k^2\left(c_s^2\dot\delta_b-\frac{1}{4}\dot\delta_\gamma\right)
    \right]+O(\tau_c^2) .
\end{equation}
This equation holds in the conformal Newtonian gauge; in the
synchronous gauge one should set $\psi=0$.  Note that $\dot\delta_b$
and $\dot\delta_\gamma$ are to be evaluated using the first of equations
(\ref{photon})--(\ref{baryon2}).  Substituting equation (\ref{sliprate})
into equation (\ref{momentum}) yields our desired equation of motion for
$\hbox{max}\,[k\tau_c,\tau_c/\tau]\ll1$.  If this condition is violated,
then one should use the explicit form of equations (\ref{baryon}) and
(\ref{baryon2}) for $\dot\theta_b$.

To obtain an equation for $\dot\theta_\gamma$ we combine the explicit
equations for $\dot\theta_\gamma$ and $\dot\theta_b$ to obtain the exact
equation
\begin{equation}
\label{thetagdot}
  \dot\theta_\gamma=-R^{-1}\left(\dot\theta_b+{\dot a\over a}\theta_b
    -c_s^2k^2\delta_b\right)+k^2\left(\frac{1}{4}\delta_\gamma-
    \sigma_\gamma\right)+{(1+R)\over R}k^2\psi
\end{equation}
in conformal Newtonian gauge; in synchronous gauge one sets $\psi=0$.
For $\dot\theta_b$ we use the tight-coupling approximation (substituting
eq. [\ref{sliprate}] into eq. [\ref{momentum}]) at early times and the
exact explicit equations (\ref{baryon}) or (\ref{baryon2}) at late times.
In practice, we switch to the explicit scheme for $\dot\theta_b$ when
$T_b=2\times10^4$ K; we switch to the explicit scheme for $\dot F_{\gamma
\,l}$ for $l>1$ and $T_b=2\times10^5$ K (at earlier times these moments
are set to zero).  We have verified that these switches occur early enough
to preserve good accuracy in the resulting photon phase space distribution.

\subsection{Recombination}

In order to compute the Thomson scattering terms in the equations of
motion for photons and baryons we need to know the free electron density
$n_e(\tau)$.  Our treatment is based on Peebles (1968; see also Peebles
1993) with the addition of helium with mass fraction $Y=0.23$.  We
summarize the procedure here.

At high temperatures, hydrogen and helium are both fully ionized.
Because of its much larger ionization potentials, helium recombines
while hydrogen is still fully ionized and the free electron density
is substantial.  Consequently, the helium recombination rates are much
larger than the expansion rate until helium recombination is essentially
complete, so that Saha ionization equilibrium is an excellent approximation.
We define the helium ionization fractions $x_1=n(\hbox{He}^+)/n(\hbox{He})$
and $x_2=n(\hbox{He}^{++})/n(\hbox{He})$, where $n(\hbox{He})$ is the
total number density of helium nuclei.  The Saha equation is
\begin{equation}
\label{saha}
   {n_e x_{n+1}\over x_n} = {2g_{n+1}\over g_n}\left(m_e k_{\rm B}T_b
     \over 2\pi\hbar^2\right)^{3/2}\,e^{-\chi_n/k_{\rm B}T_b}\,,
\end{equation}
where $n=0$ or 1 ($x_0\equiv 1-x_1-x_2$), the statistical weights for
helium are $g_0=g_1=1$ and $g_2=2$, and $\chi_n$ is the ionization
potential from $n$-times ionized helium.

When hydrogen recombines, the rapidly declining free electron density
leads to a breakdown of ionization equilibrium.  One must integrate the
appropriate kinetic equations.  Because helium recombination is
completed before hydrogen begins to recombine appreciably, it is
sufficient now to treat the helium as being fully neutral.  We define
the ionization fraction of hydrogen as $x_H=n_e/n_H$ where $n_H$
is the total number density of hydrogen nuclei.  The ionization rate equation
is (Peebles 1968; Spitzer 1978)
\begin{equation}
\label{ionrateeq}
  {d x_H\over d\tau}=aC_r\left[\beta(T_b)(1-x_H)-n_H\alpha^{(2)}(T_b)x_H^2
    \right]\ .
\end{equation}
The factor $C_r$ is discussed below.  The collisional ionization rate from
the ground state is
\begin{equation}
\label{ionrate}
  \beta(T_b)=\left(m_ek_{\rm B}T_b\over2\pi\hbar^2\right)^{3/2}
    e^{-B_1/k_{\rm B}T_b}\,\alpha^{(2)}(T_b)\ ,
\end{equation}
where $B_1=m_ee^4/(2\hbar^2)=13.6$ eV is the ground state binding energy,
and the recombination rate to excited states is
\begin{equation}
\label{recombrate}
  \alpha^{(2)}(T_b)={64\pi\over(27\pi)^{1/2}}{e^4\over m_e^2c^3}
    \left(k_{\rm B}T_b\over B_1\right)^{-1/2}\,\phi_2(T_b)\ ,\quad
    \phi_2(T_b)\approx0.448\,\ln\left(B_1\over k_{\rm B}T_b\right)\ .
\end{equation}
This expression for $\phi_2(T_b)$ is a good approximation (better
than one percent for $T_b<6000$ K).  At high temperatures this expression
underestimates $\phi_2$ but the neutral fraction is negligible
so that we make no significant error by setting $\phi_2=0$ for
$T_b>B_1/k_{\rm B}=1.58\times10^5$ K.

Recombination directly to the ground state is inhibited by the large
Lyman alpha and Lyman continuum opacities.  Net recombination must
occur either by 2-photon decay from the $2s$ level, with a rate
$\Lambda_{2s\to1s}=8.227\ {\rm s}^{-1}$, or by the cosmological redshifting
of Lyman alpha photons away from the line center.  Peebles (1968) gives
a detailed discussion of these atomic processes.  The net recombination
rate to the ground state is reduced by the fact that an atom in the
$n=2$ level may be ionized before it decays to the ground state.  The
reduction factor $C_r$ is just the ratio of the net decay rate (including
2-photon decay and Lyman alpha production at the rate allowed by redshifting
of photons out of the line) to the sum of the decay and ionization rates
from the $n=2$ level:
\begin{equation}
\label{cpeebles}
  C_r={\Lambda_\alpha+\Lambda_{2s\to1s}\over\Lambda_\alpha+\Lambda_{2s\to1s}
    +\beta^{(2)}(T_b)}\ ,
\end{equation}
where
\begin{equation}
\label{cpeebles2}
  \beta^{(2)}(T_b)=\beta(T_b)e^{+h\nu_\alpha/k_{\rm B}T_b}\ ,\quad
  \Lambda_\alpha={8\pi\dot a\over a^2\lambda_\alpha^3 n_{1s}}\ ,\quad
  \lambda_\alpha={8\pi\hbar c\over3B_1}=1.216\times10^{-5}\ {\rm cm}\ ,
\end{equation}
where $\nu_\alpha=c/\lambda_\alpha$.  For $T_b\ll 10^5$ K, it is a very
good approximation to replace the number density $n_{1s}$ of hydrogen
atoms in the $1s$ state by $(1-x_H)n_H$.

We integrate equation (\ref{ionrateeq}) using a stable and accurate
semi-implicit method with a large number of timesteps through recombination.
Since the results are independent of the perturbations, we pre-compute
the ionization history of a model and later use cubic splines interpolation
to obtain $n_e(\tau)$ (including electrons from both hydrogen and
helium) accurately during integration of the perturbation equations.
The baryon temperature and sound speed are also pre-computed for
cubic splines interpolation.

\section{Microwave Background Anisotropy}
\label{sec:cmb}

Perturbations in the photon phase space distribution correspond to
fluctuations in the cosmic microwave background radiation.  In this
section we show how to compute the anisotropy and its power spectrum
from the results of integration of the Boltzmann equation.

The photon brightness temperature perturbation $\Delta\equiv\Delta T/T$
is defined by
\begin{equation}
\label{Delta}
  f(x^i,q,n_j,\tau)=f_0\left(q\over1+\Delta\right)\ ,
\end{equation}
where $f_0(q)$ is the Bose-Einstein distribution of equation
(\ref{equil-dist}) with $\epsilon=q$.  In general the anisotropy is a
function of all phase-space variables: $\Delta=\Delta(x^i,q,n_j,\tau)$.
{}From equation (\ref{f-pert}) we obtain, to first order in $\Psi$,
\begin{equation}
  \Delta=-\left(d\ln f_0\over d\ln q\right)^{-1}\Psi\ .
\end{equation}
Because the $q$-dependence of both the gravitational source
terms and the linearized collision operator for Thomson scattering
in the Boltzmann equation for $\Psi$ (eqs. [\ref{bolt-syn}] and
[\ref{bolt-con}]) are both proportional to $d\ln f_0/d\ln q$,
$\Delta$ is independent of $q$.  In other words, the perturbed
microwave background radiation has a Planck spectrum with the
brightness temperature at a given position depending only on
the photon direction.  (Scattering by a nonlinearly perturbed medium
can, however, change the spectrum.  An example is the Sunyaev-Zel'dovich
effect occurring when the electron gas is much hotter than the
radiation.) In the absence of spectral distortions, $\Delta$ is
related very simply to the momentum-averaged phase space density
perturbation (i.e., the relative energy density perturbation):
$\Delta=\frac{1}{4}F_\gamma$.  Note that Efstathiou \& Bond (1984,
1987) and Bond (1995) define $\Delta$ to be the photon density
perturbation, 4 times our definition.

To compute the anisotropy at a given spacetime point $(x^i,\tau)$
we superpose plane-wave contributions:
\begin{equation}
\label{Delta-exp1}
  \Delta(\vec x,\hat n,\tau)=\int d^3k\,e^{i\vec k\cdot\vec x}\,
    \Delta(\vec k,\hat n,\tau)
  \equiv\int d^3k\,e^{i\vec k\cdot\vec x}\,\sum_{l=0}^\infty(-i)^l
    \,(2l+1)\,\Delta_l(\vec k,\tau)\,P_l(\hat k\cdot\hat n)\ ,
\end{equation}
where $\Delta_l=\frac{1}{4}F_{\gamma l}$.  The anisotropy at the
origin may be expanded in spherical harmonics in the usual manner:
\begin{equation}
\label{Delta-exp2}
  \Delta(\hat n)=\sum_{l=0}^\infty\sum_{m=-l}^l
    a_{lm}Y_{lm}(\hat n)\ ,\quad
  a_{lm}=(-i)^l\,4\pi\int d^3k\,Y^\ast_{lm}(\hat k)\,
    \Delta_l(\vec k,\tau)\ .
\end{equation}
The angular power spectrum of the anisotropy is defined by the
covariance matrix of these expansion coefficients:
\begin{equation}
\label{C_l-1}
  \langle a_{lm}a^\ast_{l'm'}\rangle=C_l\,\delta_{ll'}\,\delta_{mm'}\ ,
\end{equation}
where the angle brackets denote a theoretical ensemble average.
The angular power spectrum is related to the angular two-point
correlation function by
\begin{equation}
\label{C_l-2}
  C(\theta)\equiv\langle\Delta(\hat n_1)\,\Delta(\hat n_2)\rangle=
    {1\over4\pi}\sum_{l=0}^\infty (2l+1)\,C_l\,P_l(\hat n_1\cdot\hat n_2)
\end{equation}
where $\cos\theta=\hat n_1\cdot\hat n_2$.  Note that for $l>1$ the
anisotropy coefficients $\Delta_l$ are gauge-invariant.  Excluding
the monopole and dipole anisotropy, one obtains the same result for
the angular power spectrum and angular correlation function in both
synchronous and conformal Newtonian gauges.

The anisotropy coefficients $\Delta_l(\vec k,\tau)$ are random
variables with amplitudes and phases depending on the initial
perturbations.  We assume that the initial conditions can be
specified in terms of the conformal Newtonian gauge potential
$\psi$ as described in \S 7.  Because the evolution equations
for $F_{\gamma l}(\vec k,\tau)$ (and therefore $\Delta_l$) are
independent of $\hat k$, we may write
\begin{equation}
  \Delta_l(\vec k,\tau)=\psi_i(\vec k\,)\,\Delta_l(k,\tau)\ ,
\end{equation}
where $\psi_i(\vec k\,)$ is the initial perturbation and
$\Delta_l(k,\tau)$ is the solution of the Boltzmann equation
with $\psi_i=1$ (i.e., it is the photon transfer function).
Using the two-point correlation function of $\psi_i$ in Fourier
space,
\begin{equation}
  \langle\psi_i(\vec k_1)\,\psi_i(\vec k_2)\rangle=P_\psi(k)\,
    \delta_D(\vec k_1+\vec k_2)\ ,
\end{equation}
where $\delta_D$ is the Dirac delta function, we obtain the desired
result
\begin{equation}
\label{C_l}
  C_l=4\pi\int d^3k\,P_\psi(k)\,\Delta_l^2(k,\tau)\ .
\end{equation}

As an illustration of this formalism, consider the anisotropy
on large angular scales arising from the Sachs-Wolfe effect.
For isentropic perturbations on scales larger than the acoustic
horizon at recombination, the present ($\tau=\tau_0$) anisotropy
is related to the perturbation in $\psi$ at recombination by
$\Delta(\hat n,\tau_0)\approx\frac{1}{3}\psi(\vec x=-\vec n\chi,
\tau_{\rm rec})$ with $\chi=\tau_0-\tau_{\rm rec}\approx\tau_0$
(Sachs \& Wolfe 1967).  In this case the radiation transfer
function is simply $\Delta_l(k,\tau)=\frac{1}{3}j_l(k\chi)$.
Now suppose that the primeval spectrum of curvature fluctuations
is a power-law, $P_\psi(k)=A\chi^3(k\chi)^{n-4}\propto k^{n-4}$
(with $n=1$ corresponding to equal power on all scales, or the
scale-invariant Harrison-Zel'dovich spectrum).  The resulting
angular power spectrum is
\begin{equation}
C_l\approx{2^n\pi^3\over9}\,A\,{\Gamma(3-n)\,\Gamma\left(2l+n-1\over2\right)
\over\Gamma^2\left(4-n\over2\right)\,\Gamma\left(2l+5-n\over2\right)}\ .
\end{equation}
For $n=1$ this reduces to $C_l\approx(8\pi^2/9)A/[l(l+1)]$.

\section{Super-Horizon-Sized Perturbations and Initial Conditions}
\label{sec:gauge mode}

The evolution equations derived in the previous sections can be solved
numerically once the initial perturbations are specified.  We start
the integration at early times when a given $k$-mode is still outside
the horizon, i.e., $k\tau \ll 1$ where $k\tau$ is dimensionless.
(We follow common usage in referring to waves with $k\tau<1$ as being
``outside the horizon'' even though $\tau$ is more appropriately
called the comoving Hubble distance.)  The behavior of the density
fluctuations on scales larger than the horizon is gauge-dependent.
The fluctuations can appear as growing modes in one coordinate system
and as constant modes in another.  As we will show in this section,
this is exactly what occurs in the synchronous and the conformal
Newtonian gauges.

We first review the synchronous gauge behavior, which has already been
discussed by Press \& Vishniac (1980) and Wilson \& Silk (1981),
although these authors did not include neutrinos.  We are concerned
only with the radiation-dominated era since the numerical integration
for all the $k$-modes of interest will start in this era.  At this
early time, the massive neutrinos are relativistic, and the CDM
and the baryons make a negligible contribution to the total energy
density of the universe: $\bar{\rho}_{\rm total} = \bar{\rho}_\nu +
\bar{\rho}_{\gamma}$.  The expansion rate is $\dot{a}/a=\tau^{-1}$.
We can analytically extract the time-dependence of the metric and
density perturbations $h$, $\eta$, $\delta$, and $\theta$ on
super-horizon scales ($k\tau\ll1$) from equations (\ref{ein-syn}),
(\ref{massless}) and (\ref{photon}).  The large Thomson damping terms
in equations (\ref{photon}) drive the $l\ge 2$ moments of the photon
distribution function $F_{\gamma\,l}$ and the polarization function
$G_{\gamma\,l}$ to zero.  Similarly, $F_{\nu\,l}$ for $l\ge3$ can be
ignored because they are smaller than $F_{\nu\,2}$ by successive powers
of $k\tau$.  Equations (\ref{ein-syna}), (\ref{ein-sync}), (\ref{massless}),
and (\ref{photon}) then give
\begin{eqnarray}
\label{press}
  && \tau^2\ddot{h} + \tau\dot{h} + 6 [(1-R_\nu)\delta_\gamma
	+R_\nu\delta_\nu] = 0 \,, \nonumber\\
  && \dot{\delta}_\gamma + {4\over 3}\theta_\gamma + {2\over 3}\dot{h}=0\,,
	\qquad \dot{\theta}_\gamma - {1\over 4}k^2 \delta_\gamma = 0
	\,,\nonumber\\
  && \dot{\delta}_\nu + {4\over 3}\theta_\nu + {2\over 3}\dot{h}=0\,,
   	\qquad \dot{\theta}_\nu - {1\over 4}k^2
	(\delta_\nu-4\sigma_\nu) = 0 \,,\\
  && \dot{\sigma}_\nu - {2\over 15}(2\theta_\nu+\dot{h}+6\dot{\eta}) = 0
	\,, \nonumber
\end{eqnarray}
where we have defined $R_\nu\equiv\bar\rho_\nu/(\bar\rho_\gamma
+\bar\rho_\nu)$.  For $N_\nu$ flavors of neutrinos ($N_\nu=3$ in the
standard model), after electron-positron pair annihilation and before
the massive neutrinos become nonrelativistic,
$\bar\rho_\nu/\bar\rho_\gamma=(7N_\nu/8)(4/11)^{4/3}$ is a constant.

To lowest order in $k\tau$, the terms $\propto k^2$ in equations
(\ref{press}) can be dropped, and we have $\dot{\theta}_\nu=\dot
{\theta}_\gamma =0$.   Then these equations can be combined into a
single fourth-order equation for $h$:
\begin{equation}
   \tau {d^4h\over d\tau^4} + 5{d^3h\over d\tau^3}=0\,,
\end{equation}
whose four solutions are power laws: $h \propto \tau^n$ with $n=0$, 1,
2, and $-2$.  From equations (\ref{press}) we also obtain
\begin{eqnarray}
\label{modes}
	h &=& A+B(k\tau)^{-2}+C(k\tau)^2 + D(k\tau) \,, \nonumber\\
	\delta\equiv(1-R_\nu)\delta_\gamma+R_\nu\delta_\nu &=&
		-{2\over 3}B(k\tau)^{-2} - {2\over 3}C(k\tau)^2
		- {1\over 6}D(k\tau) \,, \nonumber\\
	\theta\equiv(1-R_\nu)\theta_\gamma+R_\nu\theta_\nu &=&
		-{3\over 8}Dk \,,
\end{eqnarray}
and $A$, $B$, $C$, and $D$ are arbitrary dimensionless constants.
The other metric perturbation $\eta$ can be found from equation
(\ref{ein-syna}):
\begin{equation}
	\eta = 2C + {3\over 4}D(k\tau)^{-1} \,.
\end{equation}

Press \& Vishniac (1980) derived a general expression for the time
dependence of the four eigenmodes.  They showed that of these four
modes, the first two (proportional to $A$ and $B$) are gauge modes
that can be eliminated by a suitable coordinate transformation.  The
latter two modes (proportional to $C$ and $D$) correspond to physical
modes of density perturbations on scales larger than the Hubble distance
in the radiation-dominated era.  Both physical modes appear as growing
modes in the synchronous gauge, but the $C(k\tau)^2$ mode dominates at
later times.  In fact, the mode proportional to $D$ in the
radiation-dominated era decays in the matter-dominated era (Ratra 1988).
We choose our initial conditions so that only the fastest-growing
physical mode is present (this is appropriate for perturbations created
in the early universe), in which case $\theta_\gamma=\theta_\nu=
\dot\eta=0$ to lowest order in $k\tau$.  To get nonzero starting values
we must use the full equations (\ref{press}) to obtain higher
order terms for these variables.  To get the perturbations in the
baryons we impose the condition of constant entropy per baryon.
Using all of these inputs, we obtain the leading-order behavior
of super-horizon-sized perturbations in the synchronous gauge:
\medskip
\newline {\it Synchronous gauge ---\hfil}
\begin{eqnarray}
\label{super}
	&&\delta_\gamma = -{2\over 3}C (k\tau)^2 \,, \qquad
	  \delta_c = \delta_b = {3\over 4}\delta_\nu =
		{3\over 4}\delta_\gamma \,, \nonumber\\
	&&\theta_c=0\,,\quad
	\theta_\gamma = \theta_b = -{1\over 18}C(k^4 \tau^3)\,,
	\quad \theta_\nu={23+4R_\nu \over 15+4R_\nu}
	\theta_\gamma\,,\\
	&&\sigma_\nu= {4C \over 3(15+4R_\nu)} (k\tau)^2 \,,\nonumber\\
	&&h = C(k\tau)^2 \,,\qquad
	\eta = 2C - {5+4R_\nu\over 6(15+4R_\nu)}C(k\tau)^2\,.\nonumber
\end{eqnarray}
The initial conditions for the moments $\Psi_l, l \ge 1$, of the
massive neutrino distribution can be related to $\Psi_0$ and the
variables above by equations (\ref{massive}).  To obtain the initial
$\Psi_0$, we can write the perturbed distribution function using
equation (\ref{Delta}) as
$f=f_0(q)(1+\Psi_0)=2h_{\rm P}^{-3}\{\exp[q/k(T+\Delta T)]+1\}^{-1}$,
where $\Delta T/T=\delta_\nu/4$ by the isentropic condition.  Then using
equations (\ref{massive}), we find the first three moments to be
\begin{eqnarray}
\label{psil}
 \Psi_0 &=& - {1\over 4}\delta_\nu {d\ln f_0\over d\ln q}\,,\nonumber\\
 \Psi_1 &=& - {\epsilon\over3qk}\theta_\nu {d\ln f_0\over d\ln q}\,,\\
 \Psi_2 &=& -{1\over 2} \sigma_\nu {d\ln f_0\over d\ln q}\,,\nonumber
\end{eqnarray}
where the terms proportional to $m^2_\nu a^2/q^2$ in $\Psi_2$ are dropped.
The higher moments $\Psi_l$ ($l\ge 3$) are negligible for $k\tau\ll 1$.

The initial conditions for the isentropic perturbations in the conformal
Newtonian gauge can be obtained either by solving equations (\ref{ein-con}),
(\ref{massless2}), and (\ref{photon2}), or using the transformations given
by equations (\ref{trans2}) and (\ref{deltat}) (which enables us to relate
the amplitudes in the two gauges).  We find for the growing mode
\medskip
\newline {\it Conformal Newtonian gauge ---\hfil}
\begin{eqnarray}
\label{super2}
	&&\delta_\gamma = -{40 C\over 15+4R_\nu} = -2\psi \,,\qquad
	  \delta_c = \delta_b = {3\over 4}\delta_\nu =
		{3\over 4}\delta_\gamma \,, \nonumber\\
	&&\theta_\gamma=\theta_\nu=\theta_c=\theta_b={10 C\over
		15+4R_\nu} (k^2 \tau)=\frac{1}{2}(k^2\tau)\psi \,, \\
	&&\sigma_\nu= {4C \over 3(15+4R_\nu)} (k\tau)^2=\frac{1}{15}
		(k\tau)^2\psi \,,\nonumber\\
	&&\psi = {20 C\over 15+4R_\nu} \,,\qquad
	\phi = \left( 1+{2\over 5}R_\nu \right)\psi \,.\nonumber
\end{eqnarray}
The massive neutrino moments $\Psi_l$ in this gauge are related
to $\delta_\nu$, $\theta_\nu$, and $\sigma_\nu$ by the same equations
(eq.~[\ref{psil}]) as in the synchronous gauge.
As claimed earlier, $\psi=\phi$ to zeroth order in $k\tau$ when no
neutrinos are present (i.e., $R_\nu=0$).  If we characterize the
perturbations in the conformal Newtonian gauge by the potential
$\psi$, all matter and metric variables have a very simple form
outside the horizon.  The neutrino energy fraction $R_\nu$ enters only
in the second potential $\phi$ as a result of the shear stress produced
by the free-streaming neutrinos.  Bardeen (1980) was concerned that a
large shear stress would lead to large metric perturbations in the
conformal Newtonian gauge.  We see that this does not happen for
isentropic growing-mode perturbations in which the shear stress arises
solely due to the free-streaming of relativistic collisionless particles.

We see that $\delta$ grows with time in the synchronous gauge but
remains a constant in the conformal Newtonian gauge before horizon
crossing.  Another significant difference is the larger value of the
velocity perturbations for small $k\tau$ in the conformal Newtonian gauge.
Physically, this difference arises because velocity perturbations vanish
to lowest order in the synchronous gauge because the synchronous gauge
spatial coordinates are Lagrangian coordinates for freely-falling
observers (\S 2).  The next-order velocity perturbations differ for
the neutrinos and photons because these two fluids have effectively
different equations of state: the neutrinos are collisionless while
the photons behave like a perfect fluid due to their strong coupling
to the baryons.  In the conformal Newtonian gauge, the lowest-order
velocity perturbations do not vanish because the conformal Newtonian
gauge spatial coordinates are Eulerian coordinates.  If we were to
include the next-order corrections to $\theta$ proportional to
$k^4\tau^3$, differences between the different fluid components would
appear in equations (\ref{super2}).

In the conformal Newtonian gauge, the mode proportional to $D$ in the
synchronous gauge yields $\phi\propto\psi\propto\delta\propto(k\tau)^{-3}$.
Thus, this mode corresponds to a decaying mode in the conformal Newtonian
gauge even though it yields $\delta\propto(k\tau)$ in the synchronous gauge.
The two gauge modes ($A$ and $B$ in eq. [\ref{modes}]) do not exist in
the conformal Newtonian gauge.

\section{Integration Results in Two CDM+HDM Models}
We apply the results derived in the previous sections to two spatially
flat cosmological models consisting of a mixture of CDM and HDM: (1)
the ``old'' model with a neutrino fraction of $\onu= 0.3$ and a
corresponding neutrino mass of $m_\nu = 93.13\,(\onu h^2)$ eV $= 7$ eV
(e.g. Davis et al 1992; Klypin et al 1993; Jing et al
1994; Cen \& Ostriker 1994), and (2) the $\onu= 0.2$ model
($m_\nu=4.7$ eV), which gives a better match to high-redshift
observations (Ma \& Bertschinger 1994b; Klypin et al 1995). Both models
assume $\oba=0.05$ and $H_0=50$ km s$^{-1}$ Mpc$^{-1}$.

In Fourier space, all the $\vec k$ modes in the linearized Einstein,
Boltzmann, and fluid equations evolve independently; thus the
equations can be solved for one value of $\vec k$ at a time.
Moreover, all modes with the same $k$ (the magnitude of the comoving
wavevector) obey the same evolution equations.  For purposes of
computing matter transfer functions, we integrated the equations of
motion numerically over the range 0.01 Mpc$^{-1} \leq k \leq 10$ Mpc$^{-1}$
using 31 points evenly spaced in $\log_{10} k$ with an interval of
$\Delta\log_{10} k = 0.1$.  For computing microwave background
anisotropy we used linear spacing in $k$ with 5000 points up to $k=0.5$
Mpc$^{-1}$.  The time integration was performed using the standard fifth-
and sixth-order Runge-Kutta integrator {\tt dverk} (obtained from
netlib@ornl.gov).  It began at conformal time $\tau_0= 3\times 10^{-4}$
Mpc ($z \sim 10^9$), which was chosen so that the largest $k$ (i.e.
the smallest wavelength) was well outside the horizon at the onset of
the integration.  The full integration was carried to $z=0$.
The microwave background anisotropy calculations were performed on a
Cray C-90 at the Pittsburgh Supercomputing Center, using a vectorized
code that runs at about 570 MFlops per processor.

The Einstein equations provide redundant equations for the evolution
of the metric perturbations.  In the synchronous gauge we chose to use
$a\dot h$ and $\eta$ as the primary metric perturbation variables in
the integration, and used equation (\ref{ein-synb}) and a combination
of (\ref{ein-syna}) and (\ref{ein-sync}) as the evolution equations.
In the conformal Newtonian gauge we integrated $\phi$ using equation
(\ref{ein-conb}) and obtained $\psi$ algebraically using
(\ref{ein-cond}).  In both gauges we used the time-time Einstein
equation (eqs. [\ref{ein-syna}] and [\ref{ein-cona}]) to check integration
accuracy.  In the conformal Newtonian gauge it is possible to avoid
integration of the metric perturbations altogether by combining
equations (\ref{ein-cona}) and (\ref{ein-conb}) into an algebraic
equation for $\phi$.  However, we found that this gave numerical
difficulties because the initial value of $\phi$ has to be set with
exquisite precision when $k\tau\ll1$.  We also found it necessary to
obtain the initial $\theta$'s from $\phi$ and the $\delta$s in the
combined constraint equations of (\ref{ein-cona}) and (\ref{ein-conb}).
Although the analytical expressions in equations (\ref{super2}) are
good approximations for $k\tau\ll 1$, slight deviations from the
energy-momentum constraints was found to cause numerical difficulties.

In the computations of the potential and the density fields (shown in
Figs.~\ref{fig:phi} -- \ref{fig:pow}), the photon and the massless
neutrino phase space distributions were expanded in Legendre series
(see eq. [\ref{fsubl}]) with up to 2000 $l$-values in order to guarantee
sufficient angular resolution.  In the calculations of the temperature
fluctuations (Fig.~\ref{fig:cl}), we used the criterion $l_{\rm max}
=1.5 k\tau_{\rm max}+10$, where $a(\tau_{\rm max})=1$ ($\tau_{\rm max}
\approx 12000$).
The massive neutrinos are computationally expensive due to the
momentum-dependence in equations (\ref{massive}) and (\ref{massive2}).
For all computations we performed the massive neutrino calculations on
a grid of 128 $q$-points including 50 $l$-values for every $q$.
Using our truncation schemes given by equations (\ref{truncnu}),
(\ref{truncmnu}), and (\ref{truncphot}), truncating the Boltzmann
hierarchies at $l_{\rm max}=50$ for massive neutrinos and 2000 for
the massless particles was adequate at better than the 0.1\% level.
We checked that all of our numerical approximations are adequate by
increasing the grids of $k$, $q$, and $l$ values as well as decreasing
the integration timestep.  We estimate that our final results have a
relative accuracy better than $10^{-3}$.

Figure~\ref{fig:cl} shows the angular power spectrum $C_l$ (defined in
\S 6) of the photon anisotropy in the CDM and the CDM+HDM models.  We
have assumed a scale-invariant spectrum of the primeval potential $\psi$
normalized to $P_\psi(k)=k^{-3}$.  Integrations using conformal
Newtonian and synchronous gauges agree to better than 0.1\%.  Our results
for CDM agree well with those of other workers who included accurate
numerical integrations with massless neutrinos, polarization, and helium
recombination (P. Steinhardt 1994, private communication; Hu et al. 1995).
To our knowledge, ours are the first results for the CDM+HDM models
including all of this physics and more as described in the preceding
sections.

Inclusion of massive neutrinos increases the anisotropy at large $l$
(by about 10 percent at the second and third acoustic peaks;
see Fig.~\ref{fig:cl}) because the decreased perturbation growth leads to a
$\dot\phi$ contribution to the photon energy density fluctuation
growth in equation (\ref{photon2}).  At the same time, the smaller
$\phi$ at early times with massive neutrinos leads to a slight
reduction in $C_l$ below the first acoustic peak.  The differences
between the $\Omega_\nu=0.2$ and 0.3 models are small (middle panel of
Fig.~\ref{fig:cl}).  Polarization decreases the anisotropy at high $l$
by increasing the photon shear stress $\sigma_\gamma$, leading to
increased radiative diffusion (bottom panel of Fig.~\ref{fig:cl}).

Figure~\ref{fig:phi} shows the time evolution of the metric
perturbations $\phi(k,\tau)$ and $\psi(k,\tau)$ in the conformal
Newtonian gauge for all 31 values of $k$ in the $\onu=0.2$ CDM+HDM
model.  (The metric perturbations in the synchronous gauge have no
simple physical interpretation, so we shall not bother presenting
them.)  The overall normalization was chosen arbitrarily
(corresponding to $C=-1/6$ in eqs. [\ref{super}] and [\ref{super2}]).
The difference between $\psi$ and $\phi$ in the radiation-dominated
era is due to the shear stress contributed by the relativistic
neutrinos (eq.  [\ref{super2}]) which make up a fraction
$R_\nu=0.4052$ of the total energy density.  On scales much smaller
than the horizon, $\psi$ corresponds to the Newtonian gravitational
potential and $\phi=\psi$ in the matter-dominated era.  As is well
known, the potential is constant for growing-mode density
perturbations of CDM in an Einstein-de Sitter universe.  In a mixed
dark matter model, however, the CDM density perturbation growth can be
suppressed by the lack of growth of HDM perturbations so that $\psi$
decays slowly.

The behavior of the metric perturbations can be understood as follows.
All of the 31 $k$-modes are outside the horizon at early times when
$\tau < 0.01$ Mpc.  The horizon eventually ``catches up'' and a given
$k$-mode crosses inside the horizon when $k\tau$ is about $\pi$.  The
modes with larger $k$ (i.e., shorter wavelengths) enter the horizon
earlier.  If a given $k$-mode enters the horizon during the
radiation-dominated era, the tight coupling between photons and
baryons due to Thomson scattering induces damped acoustic oscillations
in the conformal Newtonian gauge metric perturbations, which are
exhibited in Figure~\ref{fig:phi} by the modes with $k > 0.1$
Mpc$^{-1}$.  (In fact, it is not the speed-of-light horizon that sets
the scale for the oscillation and damping of the potential.  Rather,
as we show below, it is the acoustic horizon.  These horizons are
similar during the radiation-dominated era because the sound speed of
the photon-baryon fluid $c_{pb}$ is $c/\sqrt{3}$.)  The modes with $k < 0.1$
Mpc$^{-1}$ enter the horizon during the matter-dominated era and do
not oscillate acoustically because the Jeans wavenumber $k_{\rm J}=
(4\pi G\bar\rho a^2/c_{pb}^2)^{1/2}$ has then become much larger than the
wavenumbers under investigation.

We can understand the damped oscillations more quantitatively by studying
the Einstein equations (\ref{ein-con}) in the conformal Newtonian gauge.
Analytical solutions can be found for a perfect fluid with no shear
stress, in which case $\phi=\psi$.  Using $c_{pb}^2 = \delta P/\delta\rho
\approx\bar p/\bar\rho$ and $\dot{a}/a =2\tau^{-1}/(1+3c_{pb}^2)$ (from eqs.
[\ref{friedmann}] and [\ref{friedmann2}]), equations (\ref{ein-con})
can be combined to yield
\begin{equation}
   \tau^2\ddot{\phi} + {6(1+c_{pb}^2)\over
	1+3c_{pb}^2}\tau\dot{\phi} + (kc_{pb}\tau)^2\phi =0 \,,
\label{nostress}
\end{equation}
whose solutions (approximating $c_{pb}$ as a constant) are Bessel functions
with a power-law pre-factor:
\begin{equation}
	\phi_\pm = (kc_{pb}\tau)^{-\nu} J_{\pm\nu}(kc_{pb}\tau)\,,
	\qquad \nu \equiv {5+3c_{pb}^2 \over 2(1+3c_{pb}^2)} \,.
\end{equation}
The $\phi_-$ solution corresponds to the decaying mode discussed in
\S 6, so we shall ignore it.  In the radiation-dominated era, $c_{pb}^2
= \frac{1}{3}$ and $\nu=\frac{3}{2}$, so that
\begin{equation}
  \phi_+ = (kc_{pb}\tau)^{-3/2} J_{3/2}(kc_{pb}\tau)
	      \propto \cases{
		\mbox{constant}\,, & $kc_{pb}\tau \ll 1$\,, \cr
		a^{-2} \cos(kc_{pb}\tau)\,, & $kc_{pb}\tau \gg 1$\,.\cr
		}
\end{equation}
We refer to the damping for $kc_{pb}\tau\gg1$ as acoustic damping; it
corresponds to constant-amplitude acoustic oscillations in the density
contrast of the photon-baryon fluid.  The damped oscillations are apparent
in Figure~\ref{fig:phi} for $\tau < \tau_{\rm eq}$.  The analytic solution
holds of course only in the absence of neutrinos; obtaining the correct
amplitudes for $\psi$ and $\phi$ in CDM+HDM models shown in
Figure~\ref{fig:phi} required the full integration discussed in this paper.

After the universe becomes matter-dominated, acoustic damping of the
potential ceases, and the only physical process causing the potential
to change is the free-streaming damping of perturbations in the massive
neutrinos.  We shall discuss this further after examining the evolution
of the density perturbations.

Figure~\ref{fig:del} shows the evolution of the density perturbations
for the five particle species in the $\onu=0.2$ CDM+HDM model in the
two gauges from our numerical integration.  Three wavenumbers are
plotted: $k=0.01$ Mpc$^{-1}$ (Fig.~\ref{fig:del}a), $k=0.1$ Mpc$^{-1}$
(Fig.~\ref{fig:del}b), and $k=1.0$ Mpc$^{-1}$ (Fig.~\ref{fig:del}c).
Each mode is normalized with the same initial amplitude for $\phi$ as
in Figure~\ref{fig:phi}.  There are several notable features:

\noindent {\it Before horizon crossing} ---

(1) The initial amplitudes of the $\delta$'s are related by the isentropic
initial conditions: $\delta_\gamma = \delta_\nu =\delta_h = 4
\delta_b/3 = 4 \delta_c /3$.  The behavior of $\delta$ outside the
horizon is strongly gauge-dependent.  In the synchronous gauge,
Figure~\ref{fig:del} shows that all the $\delta$'s before horizon
crossing in the radiation-dominated era grow as $a^2$.  This confirms
equations (\ref{super}) since $a(\tau)\propto \tau$ at this time.
It is straightforward to show that in the matter-dominated era (for
$\Omega$=1), $\delta \propto \tau^2 \propto a$ for all modes before
horizon crossing.  In the conformal Newtonian gauge, the $\delta$'s
remain constant outside the horizon as derived in equations (\ref{super2}).

\noindent {\it After horizon crossing} ---

(2) The perturbations come into causal contact after horizon crossing
and become nearly independent of the coordinate choices.  As
Figure~\ref{fig:del} shows, $\delta_c$, $\delta_b$, and $\delta_h$ in
the two gauges are almost identical at late times.
For CDM, the $k$-modes that enter the horizon during the
radiation-dominated era behave very differently from those entering in
the matter-dominated era.  The critical scale separating the two is
the horizon distance at the epoch of radiation-matter equality
($a_{\rm eq} \sim 2\times 10^{-4} \Omega h^2$): $k_{\rm eq} =
2\pi/\tau_{\rm eq} \sim 0.1$ Mpc$^{-1}$ for our parameters.  For the
modes with $k > 0.1$, horizon crossing occurs when the energy density
of the universe is dominated by radiation; thus the fluctuations in
the CDM can not grow appreciably during this time.  For the photons
and the baryons, the important scale is the horizon size at
recombination ($a_{\rm rec}
\sim 10^{-3}$): $k_{\rm rec} = 2\pi/\tau_{\rm rec}
\sim 0.025$ Mpc$^{-1}$.  The modes with $k > 0.025$ (see
Figs.~\ref{fig:del}b and \ref{fig:del}c) enter the horizon
before recombination, so the photons (long-dashed curves) and
baryons (dash-dotted curves) oscillate acoustically while they are
coupled by Thomson scattering.  The coupling is not perfect.
The friction of the photons dragging against the baryons leads
to Silk damping (Silk 1968), which is prominent in Figure~\ref{fig:del}c
at $a\sim10^{-3.5}$.  The baryons decouple from the photons at
recombination and then fall very quickly into the potential wells
formed around the CDM, resulting in the rapid growth of $\delta_b$
in Figures~\ref{fig:del}b and \ref{fig:del}c.

(3) Neutrinos decouple from other species at $T \sim 1$ MeV and $a
\sim 10^{-10}$.  At this early time, both the massless and the eV-range
massive neutrinos behave like relativistic collisionless particles.
The massive neutrinos become non-relativistic when $3k_{\rm B}T_\nu \sim
m_\nu$, corresponding to $a_{\rm nr} \sim 10^{-4}$ for
4.7 eV neutrinos.  Close inspection of the figures at $a\approx
a_{\rm nr}$ reveals that $\delta_h$ (short-dashed curve) is
indeed making a gradual transition from the upper line for the
radiation fields to the lower line for the matter fields.  Although
the Jeans length of a fluid is not well defined for collisionless
particles such as the neutrinos, the criterion for free-streaming
damping is similar to the Jeans criterion for gravitational stability:
free-streaming is important for $k > k_{\rm fs}$, where $k^2_{\rm fs}(a)
=4\pi G\bar\rho a^2/v_{\rm med}^2$ and $v_{\rm med}$ is the median neutrino
speed.  When the neutrinos are relativistic, $v \sim 1$ and $k_{\rm fs}(a)
\propto a^{-1}$ for $a < a_{\rm eq}$.  After the neutrinos become
non-relativistic, the median neutrino speed is $v_{\rm med} = 3 k_{\rm B}
T_{0,\nu}/a\,m_\nu\ = 15 a^{-1} (m_\nu/10\,{\rm eV})^{-1}$ km s$^{-1}$.
In the matter-dominated era, we have $4\pi G\bar\rho a^2 = \frac{3}{2} H_0^2
a^{-1}$ from the Friedmann equation, and therefore
\begin{equation}
\label{kfs}
	k_{\rm fs}(a) = 8\,a^{1/2}\,\left({m_\nu \over 10
	{\rm eV}}\right) h\,{\rm Mpc}^{-1}  \,.
\end{equation}
In Figure~\ref{fig:del}a, since horizon crossing occurs when the
free-streaming effect is already unimportant, the evolution of
$\delta_h$ is very similar to that of CDM.  In Figures~\ref{fig:del}b and
\ref{fig:del}c,
however, the free streaming effect is evident and the growth of
$\delta_h$ is suppressed until $k_{\rm fs}(a)$ grows to $\sim k$.
After $k_{\rm fs}(a) > k$, $\delta_h$ can grow again and catch up to
$\delta_c$.  Since $k_{\rm fs} \propto a^{1/2}$, the larger $k$ modes
suffer more free-streaming damping and $\delta_h$ can not grow until
later times.  The damping in $\delta_h$ also affects the growth of
$\delta_c$, slowing it down more for larger $\onu$ compared to the
pure CDM model.  This effect is apparent in the $z=0$ power spectra of
the pure CDM and the two CDM+HDM models in Figure~\ref{fig:pow}.
Contrary to previous figures, the curves in Figure~\ref{fig:pow} are
normalized to the COBE-compatible rms quadrupole moment of $Q_{\rm
rms-PS}=17.6\,\mu K$ (Bennett et al. 1994).

\section{Summary}
The purpose of this paper was to present in both the synchronous and
the conformal Newtonian gauges a complete discussion of the linear
theory of scalar gravitational perturbations that is applicable to any
flat CDM, HDM or CDM+HDM model (including a possible cosmological
constant).  For historical reasons, most calculations of linear
fluctuation growth have been carried out in the synchronous gauge.
The conformal Newtonian gauge, however, offers an alternative that is
free of the gauge ambiguities and coordinate singularities associated
with the synchronous gauge.  We derived the coordinate transformation
relating the two gauges and presented in parallel in both gauges the
complete set of evolution equations: the Einstein equations for the
metric perturbations, the Boltzmann equations for the photon and
neutrino phase space distributions, and the fluid equations for CDM
and baryons.  A detailed discussion of the microwave background
anisotropy calculations was also presented.  Care was taken to include
all important higher moments of the neutrino phase space distribution
and the effects of helium recombination and photon polarization.

We solved the linear theory in the standard CDM model and two
spatially flat CDM+HDM models with $\onu=0.2$ and 0.3, assuming a
scale-invariant spectrum of isentropic primordial fluctuations.  (The
baryon fraction was taken to be $\oba=0.05$ and $H_0$=50 km s$^{-1}$
Mpc$^{-1}$.)  The evolution of the metric perturbations and the
density fields for all five particle species were presented, along
with the first accurate calculations of the photon anisotropy power
spectrum in CDM+HDM models.  We also illustrated the gauge dependence
of the density fields before horizon crossing and discussed the physical
interpretation of the results.

Interested users may obtain our programs to integrate the perturbation
equations at {\tt http://arcturus.mit.edu/cosmics}.

\acknowledgments

We thank Uros Seljak, Alan Guth, Douglas Scott, and Martin White for
helpful comments.  We particularly thank Marie Machacek for a careful
reading of the manuscript, Paul Steinhardt for pointing out the
importance of photon polarization, and Paul Bode for his assistance in
preparing a portable numerical package.  This work was supported by
NSF grant AST-9318185, NASA grants NAGW-2807 and NAG5-2816, and DOE
grant DE-AC02-76ER03069.  Supercomputing time was generously provided
by the National Center for Supercomputing Applications and the
Pittsburgh Supercomputing Center.  C.-P. Ma acknowledges
Fellowship support from the Division of Physics, Mathematics, and
Astronomy at Caltech.  E.B. would like to thank John Bahcall for his
hospitality at the Institute for Advanced Study, where part of this
work was performed.

\clearpage

\section{Figure Captions}
\begin{figure}
\epsfysize=6.5truein
\epsfbox{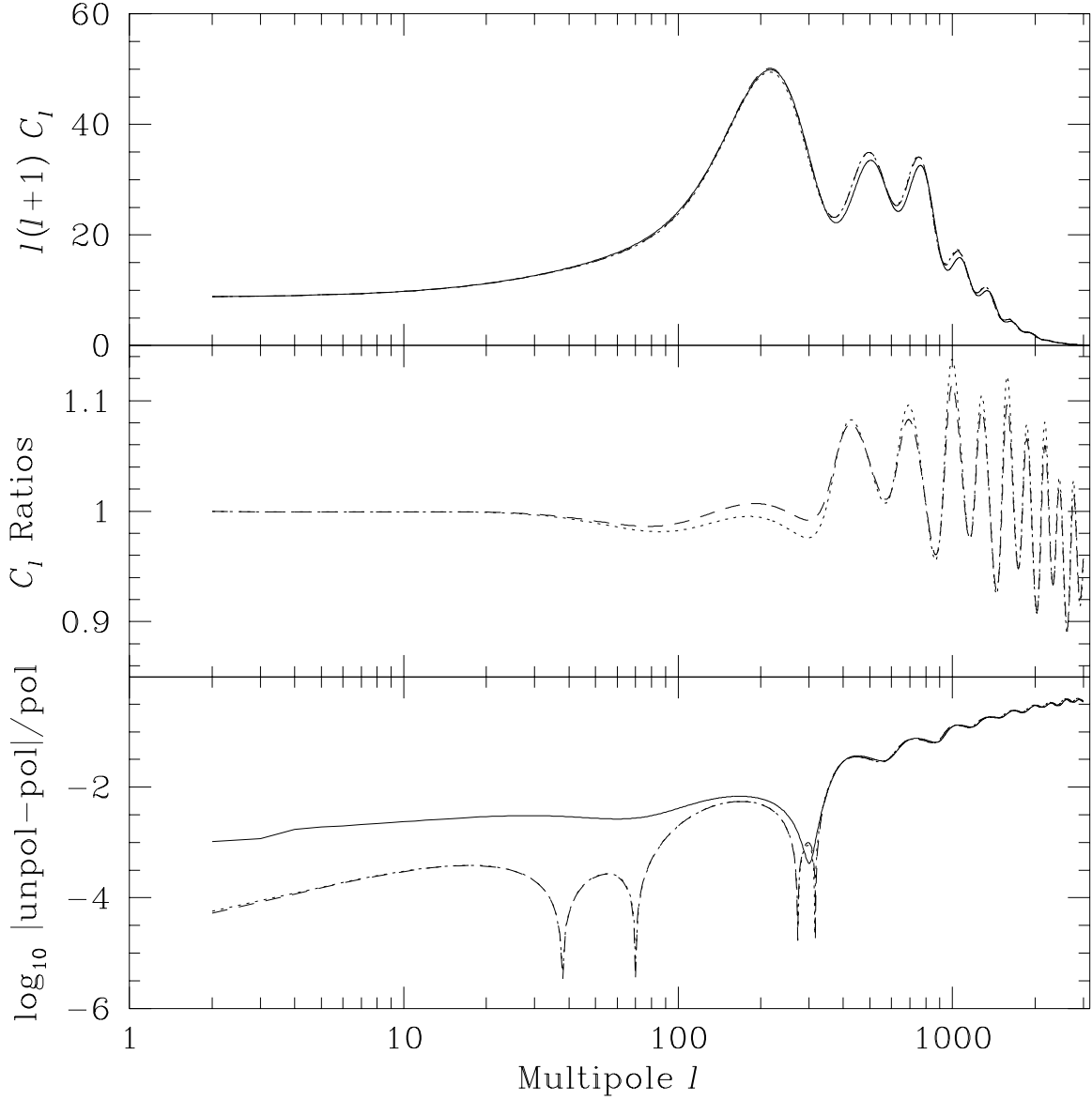}
\caption{Top: The angular power spectrum $C_l$ of the
photon anisotropy (including polarization) in the CDM (solid), the
$\onu=0.2$ CDM+HDM (dashed), and the $\onu=0.3$ CDM+HDM (dotted)
models.  Middle: The ratio of $C_l$ for the CDM+HDM models relative to
the CDM (dashed for $\onu=0.2$; dotted for $\onu=0.3$).  Bottom: The
fractional difference in $C_l$ with and without polarization for the
three models.}
\label{fig:cl}
\end{figure}

\noindent Fig.~\ref{fig:cl}:
Top: The angular power spectrum $C_l$ of the photon anisotropy
(including polarization) in the CDM (solid), the $\onu=0.2$ CDM+HDM
(dashed), and the $\onu=0.3$ CDM+HDM (dotted) models.  Middle: The
ratio of $C_l$ for the CDM+HDM models relative to the CDM (dashed for
$\onu=0.2$; dotted for $\onu=0.3$).  Bottom: The fractional difference
in $C_l$ with and without polarization for the three models.

\begin{figure}
\epsfysize=6.5truein
\epsfbox{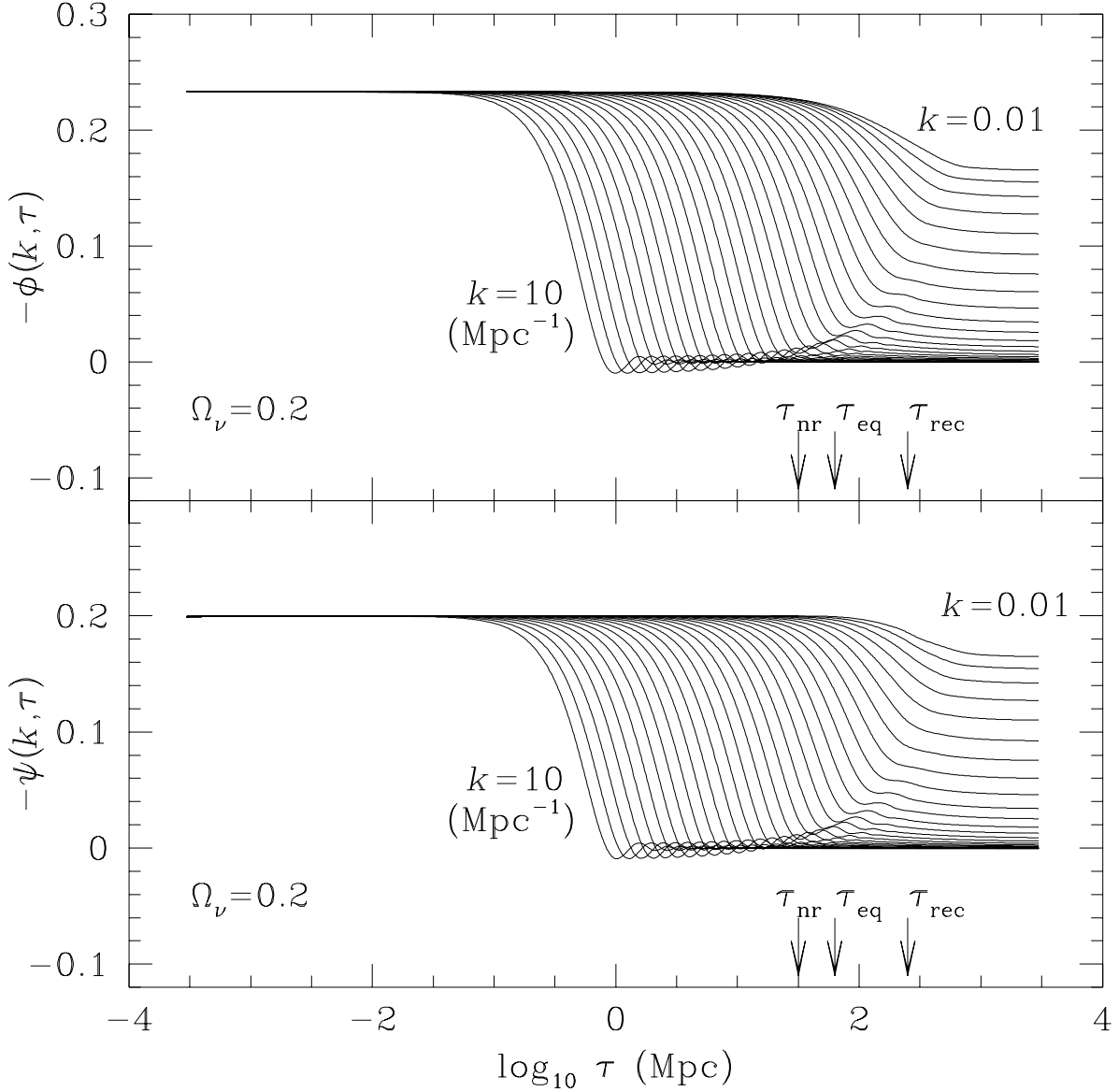}
\caption{The scalar metric perturbations $\phi(k,\tau)$ and $\psi(k,\tau)$
in the conformal Newtonian gauge in the $\onu=0.2$ CDM+HDM model.  The
31 curves from left to right correspond to 31 values of $k$ between
10.0 Mpc$^{-1}$ and 0.01 Mpc$^{-1}$.  The labels $\tau_{\rm nr}$,
$\tau_{\rm eq}$ and $\tau_{\rm rec}$ indicate, respectively, the time
the 4.7 eV neutrinos become non-relativistic, the matter-radiation
equality time, and the recombination time.}
\label{fig:phi}
\end{figure}

\noindent Fig.~\ref{fig:phi}:
The scalar metric perturbations $\phi(k,\tau)$ and $\psi(k,\tau)$ in
the conformal Newtonian gauge in the $\onu=0.2$ CDM+HDM model.  The 31
curves from left to right correspond to 31 values of $k$ between 10.0
Mpc$^{-1}$ and 0.01 Mpc$^{-1}$.  The labels $\tau_{\rm nr}$,
$\tau_{\rm eq}$ and $\tau_{\rm rec}$ indicate, respectively, the time
the 4.7 eV neutrinos become non-relativistic, the matter-radiation
equality time, and the recombination time.

\begin{figure}
\epsfysize=6.5truein
\epsfbox{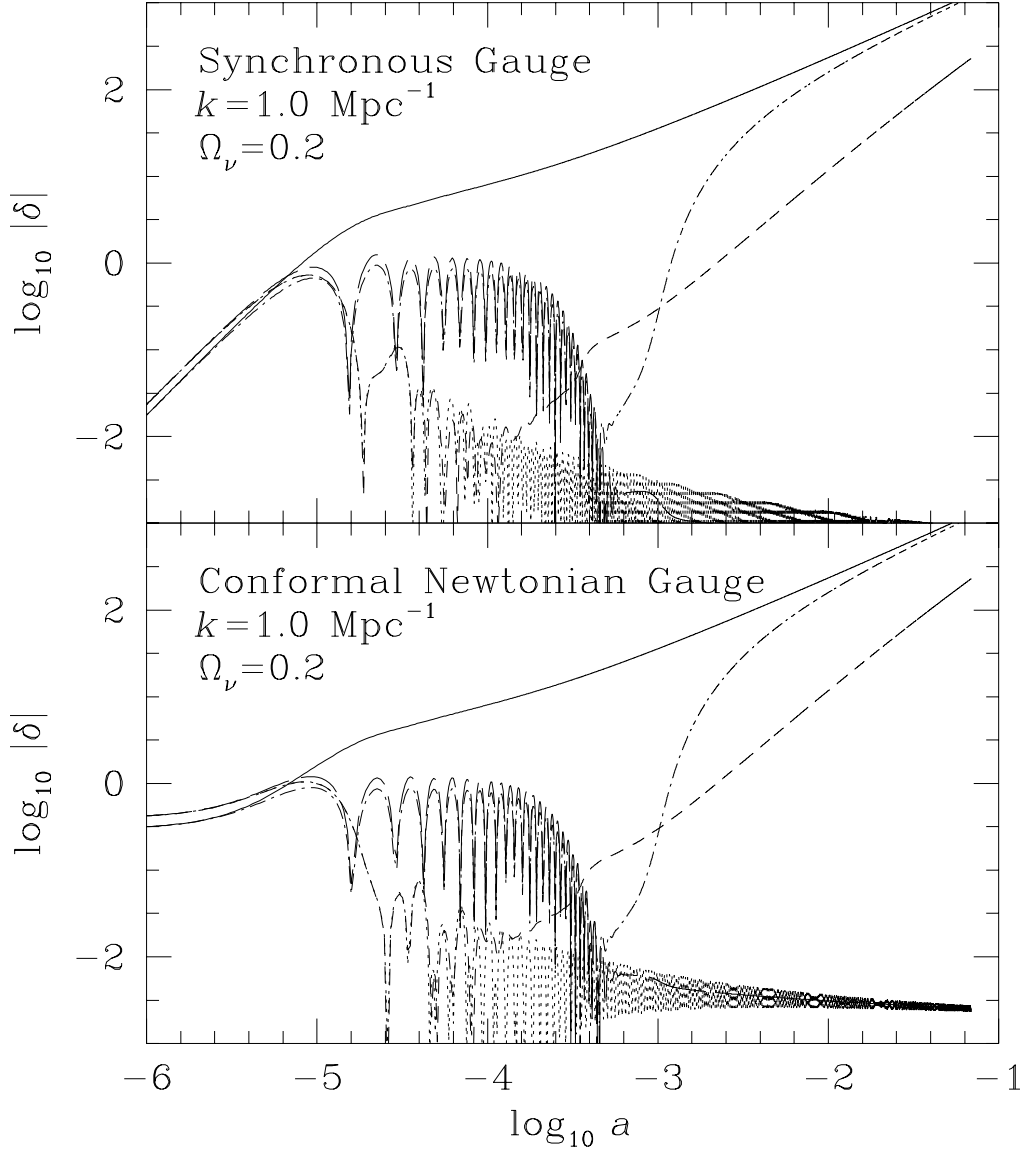}
\caption{Evolution of the density fields in the synchronous
gauge (top panels) and the conformal Newtonian gauge (bottom panels)
in the $\onu=0.2$ CDM+HDM model for 3 wavenumbers $k$= 0.01
(Fig.~\protect{\ref{fig:del}a}), 0.1 (Fig.~\protect{\ref{fig:del}b})
and 1.0 (Fig.~\protect{\ref{fig:del}c}) Mpc$^{-1}$.  	In each
figure, the five lines represent $\delta_c, \delta_b, \delta_\gamma,
\delta_\nu$ and $\delta_h$ for the CDM (solid), baryon (dash-dotted),
photon (long-dashed), massless neutrino (dotted), and massive neutrino
(short-dashed) components, respectively.}
\label{fig:del}
\end{figure}

\noindent Fig.~\ref{fig:del}:
Evolution of the density fields in the synchronous gauge (top panels)
and the conformal Newtonian gauge (bottom panels) in the $\onu=0.2$
CDM+HDM model for 3 wavenumbers $k$= 0.01
(Fig.~\protect{\ref{fig:del}a}), 0.1 (Fig.~\protect{\ref{fig:del}b})
and 1.0 (Fig.~\protect{\ref{fig:del}c}) Mpc$^{-1}$.  	In each
figure, the five lines represent $\delta_c, \delta_b, \delta_\gamma,
\delta_\nu$ and $\delta_h$ for the CDM (solid), baryon (dash-dotted),
photon (long-dashed), massless neutrino (dotted), and massive neutrino
(short-dashed) components, respectively.

\begin{figure}
\epsfysize=6.truein
\epsfbox{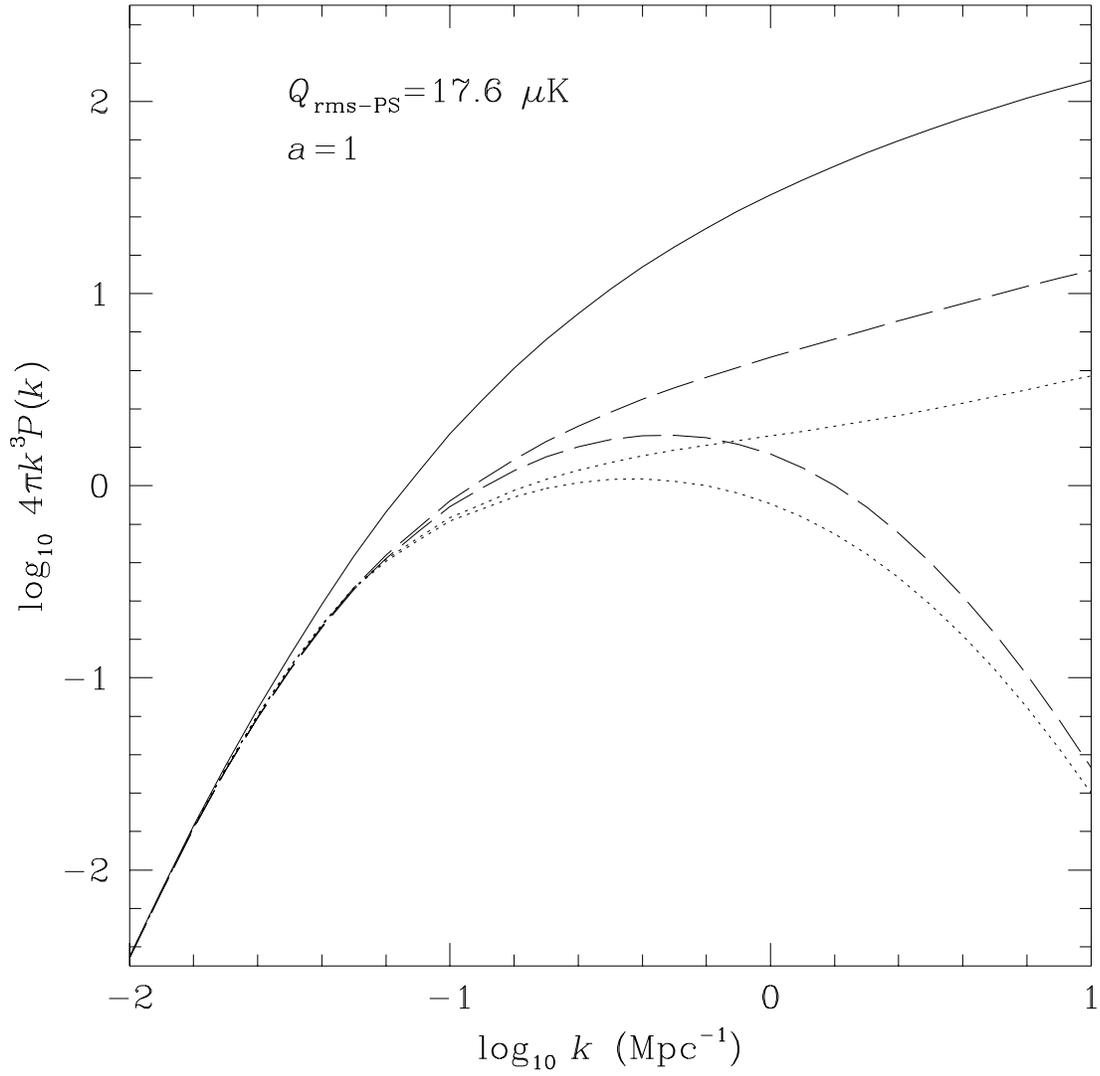}
\caption{The $z=0$ power spectra for the pure CDM (solid), the
$\onu=0.2$ CDM+HDM (dashed), and the $\onu=0.3$ CDM+HDM (dotted)
models.  In the mixed models, the upper curve is for the CDM component
and the lower one is for the HDM. The COBE-compatible quadrupole
moment of $Q_{\rm rms-PS}=17.6\,\mu\,K$ is used.}
\label{fig:pow}
\end{figure}

\noindent Fig.~\ref{fig:pow}:
The $z=0$ power spectra for the pure CDM (solid), the $\onu=0.2$
CDM+HDM (dashed), and the $\onu=0.3$ CDM+HDM (dotted) models.  In the
mixed models, the upper curve is for the CDM component and the lower
one is for the HDM.  The COBE-compatible quadrupole moment of $Q_{\rm
rms-PS}=17.6\,\mu\,K$ is used.

%%%% Insert Figure C_l

%%%% Insert Figure phi

%%%% Insert delta
\begin{figure}
\epsfysize=6.5truein
\epsfbox{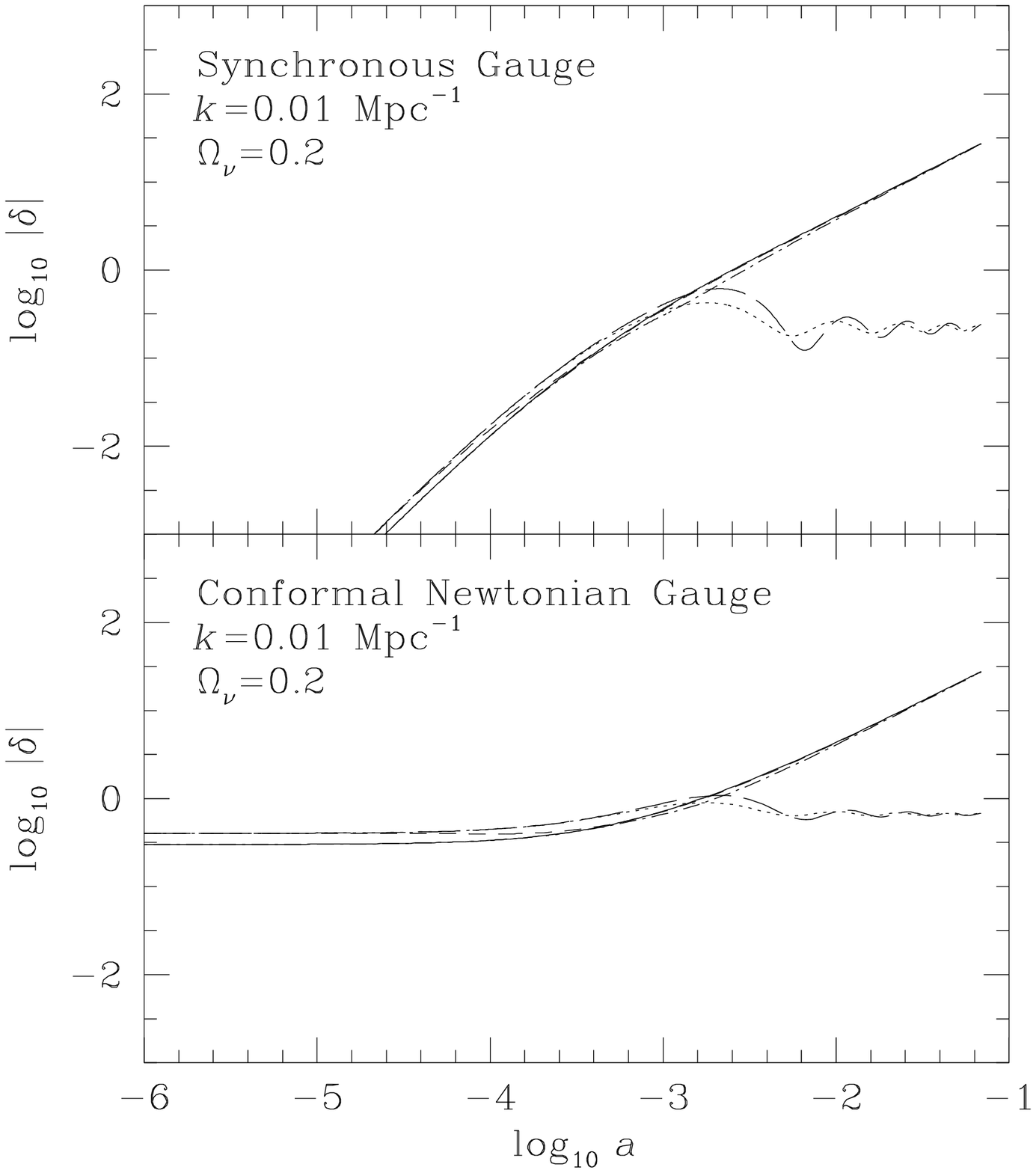}
\end{figure}
\begin{figure}
\epsfysize=6.5truein
\epsfbox{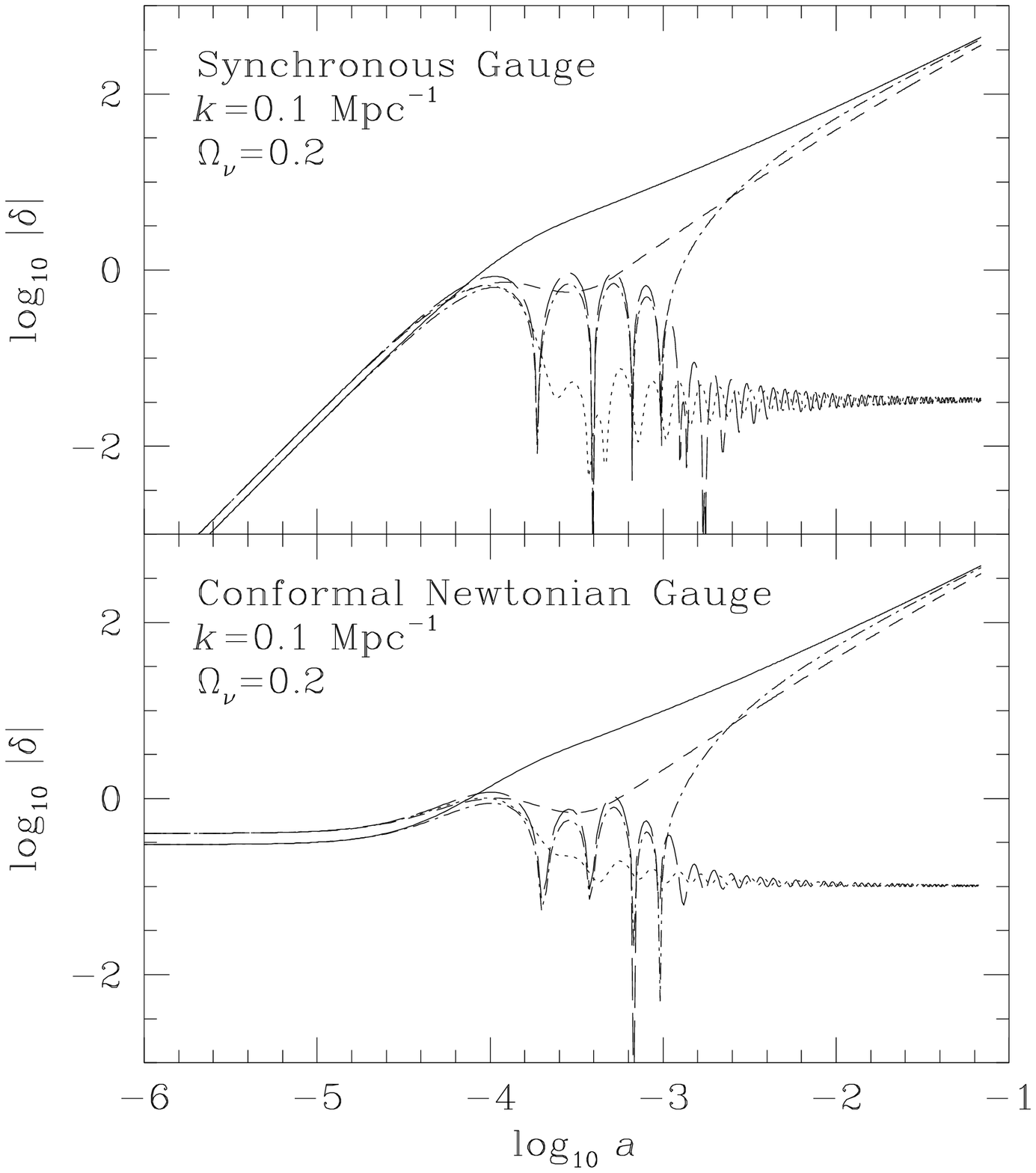}
\end{figure}

\end{document}